\documentclass[10pt,prl,aps,twocolumn,floatfix]{revtex4}
\usepackage{graphics,float}
\usepackage{amssymb}
\usepackage{subfigure}
\usepackage{amsmath,hyperref}

\usepackage{subfigure}
\usepackage [usenames,dvipsnames] {color}
\usepackage{amsbsy}
\usepackage{amsmath}
\usepackage{epsfig}
\newcommand\bea{\begin{eqnarray}}
\newcommand\eea{\end{eqnarray}}
\newcommand\beq{\begin{equation}}
\newcommand\eeq{\end{equation}}

\newcommand{\noi}{\noindent}
\newcommand{\non}{\nonumber}

\newcommand{\al}{\alpha}

\newcommand{\de}{\delta}

\newcommand{\ga}{\gamma}
\newcommand{\Ga}{\Gamma}
\newcommand{\ep}{\epsilon}

\newcommand{\ta}{\theta}

\newcommand{\da}{\dagger}

\begin{document}

\title{Steady states of a quasiperiodically driven integrable system}

\author{Sourav Nandy$^1$, Arnab Sen$^1$ and Diptiman Sen$^2$}

\affiliation{$^1$School of Physical Sciences, Indian Association for the 
Cultivation of Science, Jadavpur, Kolkata 700032, India \\
$^2$Centre for High Energy Physics, Indian Institute of Science, Bengaluru 
560012, India}

\begin{abstract}
Driven many-body quantum systems where some parameter in the Hamiltonian 
is varied quasiperiodically in time may exhibit nonequilibrium steady 
states that are qualitatively different from their periodically driven 
counterparts. Here we consider a prototypical integrable spin system, the 
spin-$1/2$ transverse field Ising model in one dimension, in a pulsed 
magnetic field. The time dependence of the field is taken to be quasiperiodic 
by choosing the pulses to be of two types that alternate according to a 
Fibonacci sequence. We show that a novel steady state emerges after an 
exponentially long time when local properties (or equivalently, reduced 
density matrices of subsystems with size much smaller than the full system) 
are considered. We use the temporal evolution of certain coarse-grained 
quantities in momentum space to understand this nonequilibrium steady state 
in more detail and show that unlike the previously known cases, this steady 
state is neither described by a periodic generalized Gibbs ensemble nor by 
an infinite temperature ensemble. Finally, we study a toy problem with a 
single two-level system driven by a Fibonacci sequence; this problem shows
how sensitive the nature of the final steady state is to the different 
parameters.
\end{abstract}

\date{\today}

\maketitle

\section{Introduction and Motivation}
\label{secI}

There has been a recent surge of interest in understanding whether
nonequilibrium steady states (NESSs) can emerge in driven many-body quantum
systems from purely unitary dynamics. This is both due to the progress in 
producing and manipulating isolated quantum systems such as ultracold quantum 
gases in experiments~\cite{Blochreview2005,Polkovnikovetalreview2011,
Goldman2014,Langenetalreview2015,Eck17} as well as the theoretical 
understanding that such states may even exhibit properties that are forbidden 
in equilibrium~\cite{ElseBN2016,KhemaniLMS2016}. Much work has focused on 
understanding NESSs in periodically driven systems, also called Floquet 
systems, where some parameter in the Hamiltonian is varied periodically in 
time (for a recent review, see Ref.~\onlinecite{MoessnerSondhireview2017}). 
The weight of evidence suggests that local properties of Floquet systems 
eventually synchronize with the period of the drive which then allows their 
description in terms of a ``periodic'' ensemble~\cite{LazaridesDM2014a, 
LazaridesDM2014b,Ponte15b,AlessioR2014}. The construction of such 
an ensemble essentially follows from Jaynes' principle of maximum 
entropy~\cite{Jaynes1957a,Jaynes1957b} where the appropriate constants of 
motion are taken to be {\it all} local quantities that are stroboscopically 
conserved. In a generic many-body interacting system, it is expected that 
{\it no} such quantity exists~\cite{NandkishoreH2015} since the Hamiltonian 
is not conserved any more, and thus the system should eventually reach an 
infinite temperature steady state as far as local properties are 
concerned~\cite{Russomanno12,Bukov15,Ponte15a,Eckardt15}.

Integrable spin systems (such as the one-dimensional transverse field Ising 
model or the two-dimensional Kitaev model) that are reducible to free fermions 
via a non-local Jordan Wigner transformation~\cite{Subirbook,Duttabook,
FengZX2007,ChenN2008} provide an important exception to this rule. Here, it is 
still possible to define an extensive number of local quantities that are 
conserved stroboscopically even when the Hamiltonian of the system is 
periodically driven in time. Applying Jaynes' principle then leads to a 
periodic generalized Gibbs ensemble (p-GGE)~\cite{KollarWE2011,LazaridesDM2014a}
which is completely different from the infinite temperature ensemble (ITE)
expected for generic systems. Several previous works have focused on various 
aspects of such periodically driven integrable systems like topological
transitions~\cite{KitagawaBRD2010}, defect 
generation~\cite{Russomanno12,Mukherjee08}, dynamical freezing~\cite{Das10}, 
work statistics~\cite{DuttaDS2015} and the entanglement generation and nature 
of approach to the final NESS~\cite{SenNS2016,NandySS2018}. 
However, the nature of possible NESSs when such integrable systems are 
continually driven without any periodic structure in time is still an open 
issue. Naively, one might suspect that in the absence of any periodicity in 
time, it is not possible to construct any local conserved quantities and hence 
the system heats up to an ITE in spite of its integrability. 

Such aperiodic drives can arise in several ways when some parameter, $g$, of 
the Hamiltonian of the system is varied in time. For concreteness, we will 
consider the one-dimensional (1D) transverse field Ising model (TFIM) in 
this work where the time-dependent Hamiltonian is defined as
\bea H(t) ~=~ - ~\sum_{j=1}^L ~[g(t)s_j^x ~+~ s_j^z s^z_{j+1}], 
\label{1dTFIM} \eea
where $s_j^{x,y,z}$ are Pauli matrices describing a spin-1/2 object at site 
$j$, $L$ denotes the number of spins in 
the chain, and periodic boundary conditions are assumed in space. Here $g(t)$ 
represents a time-dependent transverse magnetic field which drives the 
system continually. For a Floquet problem, $g(t)$ is a strictly periodic 
function that satisfies $g(t) = g(t+nT)$ where $n$ is any integer and 
$T=2\pi/\omega$ is the period of the drive with $\omega$ being the drive 
frequency. Now suppose that $g(t)$ instead follows one of the two relations 
below:
\begin{subequations}
\begin{align}
g(t) &=~ \cos(\omega t) ~+~ \de g(t) \\
g(t) &=~ \cos(\omega t) ~+~ \cos( \al \omega t),
\end{align}
\label{perturbed_g} \end{subequations} 
where $\omega$ is a given drive frequency, $\de g(t)$ is a 
rapidly fluctuating white noise with zero average, and $\al$ is any 
irrational number like the golden ratio $\varphi=(1+\sqrt{5})/2$. 
Eq.~\eqref{perturbed_g} (a) represents the case where $g(t)$ is a periodic 
function perturbed by a random noise while Eq.~\eqref{perturbed_g} (b) 
represents a case that is neither periodic nor random, 
but quasiperiodic in time. Refs.~\onlinecite{MarinoS2012,RooszJI2016} 
considered a 1D TFIM in a noisy magnetic field (Eq.~\eqref{1dTFIM}), though not 
periodically driven on average (so that $g(t) = \de g(t)$), and found that the
asymptotic steady state is indeed an ITE. Recently, we have considered a case 
very similar to Eq.~\eqref{perturbed_g} (a) in Ref.~\onlinecite{NandySS2017} 
and showed that even when $g(t)$ is periodic on average and $\de g(t)$ can 
be considered to be a small perturbation, the system eventually heats up to an 
ITE in a diffusive manner after initially being in a prethermal regime where 
it is close to the p-GGE for the perfectly periodic situation. In
Ref.~\onlinecite{NandySS2017} we have also shown that not {\it all} aperiodic
drives necessarily lead to an ITE by considering a case where the perturbing 
``noise'' is not random but scale-invariant in time and showing that the 
asymptotic steady state is then described by a new geometric generalized Gibbs 
ensemble that arises due to an {\it emergent} time periodicity in the unitary 
dynamics of the driven 1D TFIM.

Our main motivation in this work is to further understand whether well-defined
NESSs that are non-ITEs can emerge for aperiodically driven integrable models 
where the driving protocol is neither random nor periodic but quasiperiodic in 
time. We will consider a simpler case analogous to Eq.~\eqref{perturbed_g} (b)
where the spectrum of the driving function shows pronounced peaks in 
Fourier space at frequencies that are incommensurate multiples of each other
(Fig.~\ref{fig1}). Using appropriate coarse-grained quantities in momentum 
space that fully determine the reduced density matrix of a subsystem of $l$ 
consecutive spins for any $l \ll L$~\cite{LaiY2015,NandySDD2016} when 
$L \gg 1$, we will show that a well-defined NESS does emerge for the 
quasiperiodic drive protocol considered here. The behaviour of these 
coarse-grained quantities further shows that this NESS is {\it qualitatively 
different} from the previously known NESSs for the continually driven 1D TFIM.

The rest of the paper is organized in the following manner. In 
Sec.~\ref{sec:IIa}, we discuss the pseudospin representation for the 1D 
TFIM that allows the many-body wave function to be denoted in terms of points 
on Bloch spheres in momentum space. In Secs.~\ref{sec:IIb} and~\ref{sec:IIc}, 
we introduce a reduced density matrix and certain coarse-grained quantities in 
momentum space which fully determine the reduced density matrix of any 
subsystem $l \ll L$. In Sec.~\ref{sec:IId} we specify the quasiperiodic 
drive protocol that we adopt in the rest of the paper. Sec.~\ref{sec:IIe} 
discusses the trajectory on the Bloch sphere for a p-GGE and an ITE. In 
Secs.~\ref{sec:IIIa} - \ref{sec:IIIc}, we present the results for some local 
quantities and for the coarse-grained quantities for quasiperiodic driving, 
and we show that a NESS exists at exponentially late times. In 
Sec.~\ref{sec:IIId}, we present an invariant which had been shown long ago to 
be very useful for studying Fibonacci sequences of $2 \times 2$ 
matrices~\cite{KohmotoKT1983,Sutherland1986}. In Sec.~\ref{sec:IV}, we study a 
toy version of the full problem and show interesting ``geometric'' transitions 
as a function of the various parameters of the problem. Finally, we summarize 
our results and conclude with some future directions in Sec.~\ref{sec:V}.

\section{Some preliminaries}
\label{sec:II}

\subsection{Pseudospin representation of dynamics of the 1D TFIM}
\label{sec:IIa}

The 1D TFIM (Eq.~\eqref{1dTFIM}) can be solved by introducing the standard 
Jordan-Wigner transformation of spin-1/2's to spinless fermions~\cite{Lieb61}
as
\bea s_n^x &=& 1 ~-~ 2c^\da_n c_n, \non \\
s_n^z &=& - ~(c_n+c_n^\da) ~\prod_{m<n} ~(1 ~-~ 2c_m^\da c_m). \label{jw1} \eea
We write $H$ in Eq.~\eqref{1dTFIM} in terms of these fermion operators and 
focus henceforth on the sector with an even number of fermions. We take $L$ 
to be even with antiperiodic boundary conditions for the fermions 
since that corresponds to periodic boundary conditions for the spins. 
This reduces the problem to free fermions with the Hamiltonian
\bea H &=& - ~\sum_{n=1}^L ~\bigl[ g(t) ~-~ 2g(t) ~c^\da_n c_n ~+~ c^\da_n 
c_{n+1} ~+~ c^\da_{n+1}c_n \non \\
&& ~~~~~~~~~~+~ c_{n+1} c_n ~+~ c_n^\da c_{n+1}^\da \bigr]. \label{jw2} \eea
To exploit the translational symmetry of the system, we go to $k$ space 
by defining 
\beq c_k ~=~ \frac{e^{i\pi/4}}{\sqrt{L}} ~\sum_{n=1}^L ~e^{-ikn} ~c_n, 
\label{jw3} \eeq
where the momentum $k=2\pi m/L$ with $m=-(L-1)/2, \cdots, -1/2,1/2,\cdots,
(L-1)/2$ for even $L$. Rewriting the Hamiltonian in terms of $c_k$ and 
$c^\da_k$, we get
\bea H &=& \sum_k ~\bigl[2(g(t) ~-~ \cos k) ~c_k^\da c_k ~+~ \sin k ~c_{-k}c_k 
\non \\ 
&& ~~~~~~+ ~\sin k ~c^\da_k c^\da_{-k} ~-~ g(t) \bigr]. \label{jw4} \eea
We now introduce a ``pseudospin representation'' $\sigma_k$ (that is different 
from the $s_j$ matrices in Eq.~\eqref{1dTFIM}), where $|\uparrow\rangle_k = 
|k,-k \rangle = c^\da_k c^\da_{-k}|0 \rangle$ and $|\downarrow
\rangle_k = |0 \rangle$~\cite{KolodrubetzCH2012}, where $|0 \rangle$ 
represents the vacuum of the $c$ fermions. In this basis, we can write 
Eq.~\eqref{jw4} as 
\beq H ~=~ \sum_{k>0} ~\left( \begin{array}{cc}
c_k^\dagger & c_{-k} \end{array} \right) ~H_k \left( \begin{array}{c}
c_k \\ c_{-k}^\dagger \end{array} \right), \eeq 
where 
\bea H_k ~=~ 2 ~(g(t)-\cos k) ~\sigma_k^z ~+~ 2 \sin k ~\sigma_k^x. 
\label{pseudospinH} \eea 
The pseudospin state $|\psi_k\rangle$ for each $k$ mode then evolves 
independently due to its own time-dependent ``pseudo-magnetic field'' given by 
Eq.~\eqref{pseudospinH}. The many-body wave function $|\psi(t) \rangle$ 
can then be expressed as 
\bea |\psi(t) \rangle &=& \otimes_{k>0} ~|\psi_k(t) \rangle, \non \\
|\psi_k(t) \rangle &=& u_k(t)|\uparrow \rangle_k ~+~ v_k(t)|\downarrow 
\rangle_k. \label{manybodyw} \eea
Eq.~\eqref{manybodyw} implies that
specifying the column vector $(u_k(t),v_k(t))^T$ (where the superscript 
$T$ denotes the transpose of the row) for each allowed value of $k$ for a 
finite $L$ completely specifies the wave function $|\psi(t) \rangle$.
The state $(u_k(t),v_k(t))^T$ can equivalently be 
represented as a point that evolves in time due to Eq.~\eqref{pseudospinH} 
on the corresponding Bloch sphere by using 
\bea |\psi_k(t) \rangle &=& \cos \left(\frac{\theta_k(t)}{2} \right) |\uparrow 
\rangle_k ~+~ e^{i \phi_k(t)}\sin \left(\frac{\theta_k(t)}{2} \right) 
|\downarrow \rangle_k, \non \\
&& \label{blochsphere} \eea
where $0 \leq \theta_k(t) \leq \pi$ and $0 \leq \phi_k(t) \leq 2\pi$. In the 
rest of this paper, we will take $(u_k,v_k)^T= (0,1)^T$ to be the initial 
state at each $k$. This corresponds to the system being initially prepared in 
a pure state where all the (physical) spins are $s^x=+1$ (this corresponds to 
the ground state of the system when $g \to \infty$). 

\subsection{Reduced density matrix for $l$ adjacent spins}
\label{sec:IIb}

Given the many-body wave function $|\psi(t) \rangle$, all local properties 
within a subsystem of $l$ adjacent spins can be understood by considering 
the reduced density matrix $\rho_l(t)$ given by 
\bea \rho_l(t) = \mathrm{Tr}_{L-l}(\rho(t)), \label{RDM} \eea
where $\rho(t) = |\psi(t) \rangle \langle \psi(t)|$ is the full density matrix 
and ${\rm Tr}_{L-l}$ indicates that 
all the $L-l$ degrees of freedom outside the subsystem have been integrated
out. Though the full density matrix $\rho(t)$ is a pure density matrix 
because of the unitary nature of the dynamics, the reduced density matrix 
$\rho_l(t)$ is typically mixed since $|\psi(t) \rangle$ gets more entangled 
as the system is driven. When $|\psi(t)\rangle$ has the form given in 
Eq.~\eqref{manybodyw}, the reduced matrix of $l$ adjacent spins (the 
correspondence is not straightforward for a subsystem with non-adjacent spins)
is determined in terms of the $c$ fermion correlation functions at these $l$ 
sites.~\cite{VidalLRK2003,Peschel2003} For free fermions, since all 
higher-point correlations are defined in terms of the two-point correlations 
from Wick's theorem, the reduced density matrix is fully determined from 
two $l\times l$ matrices~\cite{VidalLRK2003,Peschel2003}, $\mathbf{C}$ and 
$\mathbf{F}$, whose elements are defined as
\bea C_{ij} &=& \langle c_i^\da c_j \rangle_t ~=~ \frac{2}{L}\sum_{k>0}
|u_k(t)|^2 \cos[k(i-j)], \non \\
F_{ij} &=& \langle c_i^\da c_j^\da \rangle_t ~=~ \frac{2}{L}
\sum_{k>0}u^*_k(t) v_k(t) \sin[k(i-j)], \label{peschel1} \eea
where $i$, $j$ refer to two sites that belong to the subsystem. 
Using Eq.~\eqref{peschel1}, we construct the following $2l \times 2l$ matrix 
\bea {\mathcal C}_t(l) &=& \left( \begin{array}{cc} \mathbf{I-C} &
\mathbf{F}^{\da} \\ \mathbf{F} & \mathbf{C } \end{array} \right).
\label{peschel2} \eea
The eigenvalues and eigenvectors of $\mathcal{C}_t(l)$ completely determine 
the reduced density matrix $\rho_l(t)$ of the subsystem. For example, the 
entanglement entropy equals
\bea S_{\mathrm{ent}}(l) = -\mathrm{Tr}[\rho_l(t) \log (\rho_l(t))] = -
\sum_{k=1}^{2l}p_k \log(p_k), \label{entanglement} \eea
where $p_k$ is the $k$-th eigenvalue of ${\mathcal C}_t(l)$.

\subsection{Coarse-graining in $k$ space}
\label{sec:IIc}

It has been noted previously~\cite{LaiY2015,NandySDD2016,NandySS2017} 
that while the entire wave function $|\psi(t) \rangle$ requires 
specifying $(u_k(t),v_k(t))^T$ at $k=2\pi m/L$ for $k=1/2,3/2,\cdots,(L-1)/2$ 
(Eq.~\eqref{manybodyw}), $\mathcal{C}_t(l)$ and therefore the reduced density 
matrix $\rho_l(t)$ for any $l \ll L$ depends {\it only} on certain 
coarse-grained variables defined in $k$ space as follows:
\bea (|u_k(t)|^2)_c &=& \frac{1}{N_c} \sum_{k \in k_{\mathrm{cell}}} 
|u_k(t)|^2, \non \\
(u_k^*(t)v_k(t))_c &=& \frac{1}{N_c} \sum_{k \in k_{\mathrm{cell}}}
u^*_k(t)v_k (t). \label{coarsegrain} \eea
For a system with $L \gg 1$, the above variables are defined using $N_c (\gg
1)$ consecutive $k$ modes that lie within a cell (denoted by 
$k_{\mathrm{cell}}$) which has an average momentum that we denote by $k_c$ and
a size $\de k$ such that
\bea 1/L \ll \de k \ll 1/l. \label{bzcondition} \eea
With this condition on $\de k$, it is easy to see that for a subsystem with 
$l$ adjacent spins, the sinusoidal factors in Eq.~\eqref{coarsegrain}
in each momentum cell can be replaced by 
\bea \cos[k(i-j)] &\simeq& \cos[k_c(i-j)], 
\non \\
\sin[k(i-j)] &\simeq& \sin[k_c(i-j)], 
\label{simplestep} \eea 
for all values of $|i-j|$ lying in the range $[0,l]$.
The sum over the $k$ modes in Eq.~\eqref{coarsegrain} can then be carried 
out in two steps: first, summing over the consecutive momenta in a single 
coarse-grained cell, and then summing over all the different momentum cells. 
This immediately gives 
\bea C_{ij} &\simeq& \frac{1}{\mathcal{N}_{\mathrm{cell}}} ~\sum_{k_c} ~
(|u_k(t)|^2)_c \cos[k_c(i-j)], \non \\
F_{ij} &\simeq& \frac{1}{\mathcal{N}_{\mathrm{cell}}} ~\sum_{k_c} ~
(u^*_k(t)v_k(t))_c \sin[k_c(i-j)], \label{} \eea
where $\mathcal{N}_{\mathrm{cell}}$ is the number of momentum cells after the
coarse-graining in $k$ space. To take the thermodynamic limit of these 
coarse-grained quantities (Eq.~\eqref{coarsegrain}), we keep 
$\mathcal{N}_{\mathrm{cell}}$ fixed and take $L \to \infty$. 

Since each $(u_k(t),v_k(t))^T$ can be represented by a point
$(\theta_k(t),\phi_k(t))$ on the Bloch sphere, the behaviour of the 
coarse-grained quantities in Eq.~\eqref{coarsegrain} depend on the simultaneous 
positions of $(\theta_k(t),\phi_k(t))$ on the surface of a unit sphere 
of all the momentum modes that lie in a coarse-grained cell. Since their
number $N_c \gg 1$, it is useful to instead define a density distribution
of these $N_c$ points on the unit sphere, denoted by 
$P_t(\cos \theta_{k_c},\phi_{k_c})$, and study its evolution for a momentum
cell as a function of $t$. If a NESS is reached for a subsystem of size $l$,
then it necessarily implies that the coarse-grained quantities defined in
Eq.~\eqref{coarsegrain} for any momentum cell that respects 
Eq.~\eqref{bzcondition} must reach a well-defined steady state as $L \to
\infty$. Similarly, $P_t(\cos \theta_{k_c},\phi_{k_c})$ will
also have a well-defined large $t$ limit, which we denote by $P_\infty(\cos
\theta_{k_c},\phi_{k_c})$, for any such momentum cell. The functions 
$P_\infty(\cos\theta_{k_c},\phi_{k_c})$, and the steady state values of 
$(|u_k(t)|^2)_c$ and $(u_k^*(t)v_k(t))_c$ also characterize the precise nature 
of the NESS. However, no such well-defined large $t$ limit can exist for a 
single $k$ mode since $(u_k(t), v_k(t))$ will continue to display Rabi-type 
oscillations as the pseudospin $\sigma_k$ is acted upon by a 
time-dependent pseudo-magnetic field (Eq.~\eqref{pseudospinH}).

\subsection{Details of the quasiperiodic drive protocol}
\label{sec:IId}

For a periodic protocol with time period $T$, it is sufficient to evolve 
the state stroboscopically to find the p-GGE, 
\bea |\psi(nT) \rangle = U(T)^n |\psi(0) \rangle, \label{periodic} \eea
where $U(T)$ is the time evolution operator for one time period. 
Eq.~\eqref{periodic} thus generates a discrete quantum map indexed by $n$ and 
the steady state is obtained when $n \to \infty$. It is rather difficult to 
numerically simulate a quasiperiodic drive protocol composed of two 
incommensurate frequencies (Eq.~\eqref{perturbed_g}(b)); we therefore
adopt a simpler function $g(t)$ which shares the feature of having sharp 
peaks in Fourier space at frequencies that are incommensurate multiples 
of each other. We consider a reference function $g_{\mathrm{ref}}(t)$ which we 
take to be a square pulse in time for mathematical convenience:
\bea g_{\mathrm{ref}}(t) &=& g_i ~~~ \mathrm{for} ~~~~ 0 ~\leq~ t ~\leq~ T/2, 
\non \\
&=& g_f ~~~ \mathrm{for} ~~~ T/2 ~\leq t~ \leq~ T.
\label{squarepulse} \eea
The corresponding unitary time evolution operator for the mode with 
momentum $k$ can be written as
\bea U_k(T) = \exp \left(-iH_k (g_f) \frac{T}{2} \right) \exp \left( -iH_k(g_i)
\frac{T}{2}\right), \label{Uk} \eea
where $H_k$ is of the form given in Eq.~\eqref{pseudospinH}. Next, we define 
two 
types of pulses from Eq.~\eqref{squarepulse} by simply replacing $T$ by $T+dT$ 
and $T-dT$. The corresponding time evolution operators $U_k(T \pm dT)$ can then
be obtained from Eq.~\eqref{Uk} by replacing $T \to T \pm dT$. We now consider 
a different drive protocol where $g(t)$ is obtained from these two types of 
pulses which are made to alternate according to a Fibonacci sequence, which 
is a well-known quasiperiodic sequence. It is useful to note here 
that the protocol described above involving two time periods $T + dT$ and 
$T-dT$ is mathematically equivalent to a protocol in which the time period
$T$ is kept fixed but the Hamiltonian is scaled globally by factors of 
$1+dT/T$ and $1-dT/T$, respectively. 

The discrete quantum map that we study for this $g(t)$ is as follows:
\begin{widetext}
\bea |\psi_k(n) \rangle ~=~ U_k(T+ \tau_n dT) U_k(T + \tau_{n-1}dT) \cdots 
U(T + \tau_1 dT) |\psi_k(0) \rangle ~=~ {\cal T} ~\prod_{i=1}^n ~U_k(T+\tau_i
dT) |\psi_k(0) \rangle, \label{perturbedFloquet} \eea
\end{widetext}
where the sequence ${\tau_i}=\tau_1, \tau_2, \tau_3, \cdots$ (with each 
$\tau_i$ being equal to either $+1$ or $-1$) is the same for all the allowed 
$k$ modes at a finite $L$, and $\cal T$ denotes time ordering. 

The sequence ${\tau_i}$ determines the characteristics of the 
noise~\cite{NandySS2017} added to the periodic problem and its strength is 
controlled by $dT/T$. For example, if the sequence ${\tau_i}$ is taken to be 
any typical realization of a random process where each element is chosen to be 
$\pm 1$ randomly and independently with probability $1/2$, then it mimics the 
case shown in Eq.~\eqref{perturbed_g}(a). As an example of a Floquet system 
perturbed with a scale-invariant noise, the well-known Thue-Morse 
sequence~\cite{Thue1906,Morse1921a,Morse1921b} was studied in 
Ref.~\onlinecite{NandySS2017} since the sequence is self-similar by 
construction. 

Here, we take the sequence ${\tau_i}$ to be given by the Fibonacci 
sequence~\cite{Schroederbook1991}, which is perhaps the best-known example of 
a quasiperiodic sequence; this is an infinite sequence of $\tau_i=\pm 1$ that 
is obtained by starting with $\tau_1=+1$ at level 1, and $\tau_1=+1, ~
\tau_2=-1$ at level 2. The elements at level $m$ are then obtained recursively 
by following the elements at level $m-1$ with those at level $m-2$.
The first few steps of this recursive procedure yield
\begin{widetext}
\bea 
m=1:&& ~~ \tau_1 ~=~ +1 \non \\
m=2:&& ~~ \tau_1,\tau_2 ~=~ +1,-1 \non \\
m=3:&& ~~ \tau_1, \tau_2, \tau_3 ~=~ +1,-1,+1 \non \\
m=4:&& ~~ \tau_1, \cdots, \tau_5 ~=~ +1,-1,+1,+1,-1 \non \\
m=5:&& ~~ \tau_1, \cdots, \tau_{8} ~=~ +1,-1,+1,+1,-1,+1,-1,+1 \non \\
m=6:&& ~~ \tau_1, \cdots, \tau_{13} ~=~ +1,-1,+1,+1,-1,+1,-1,+1,+1,-1,+1,+1,-1 
\non \\
&& \vdots \label{recursiveF} \eea
\end{widetext}

The number of elements at each recursion level $m$ increases in accordance to 
the Fibonacci numbers ($1,2,3,5,8,13,\cdots$). Since the ratio of consecutive 
Fibonacci numbers is known to approach the golden ratio $\varphi =(1+
\sqrt{5})/2$ asymptotically, the number of elements $n$ generated at 
level $m$ increases as $n \sim \varphi^m$ for large $m$. Henceforth, we use $n$ 
in place of time $t$ since we will study the discrete quantum map defined in 
Eq.~\eqref{perturbedFloquet} and use the recursion level $m$ as a shorthand for 
denoting $n$ drives where $n=F_m$, where the $m$-th Fibonacci number $F_m$ is 
defined by the standard rule $F_m=F_{m-1}+F_{m-2}$ (for $m \ge 3$) with 
$F_1=1$ and $F_2=2$. The case of a single
spin-$1/2$ subjected to such a pulsed magnetic field following the Fibonacci
sequence has been studied by Sutherland~\cite{Sutherland1986}.

\begin{figure}
{\includegraphics[width=\hsize]{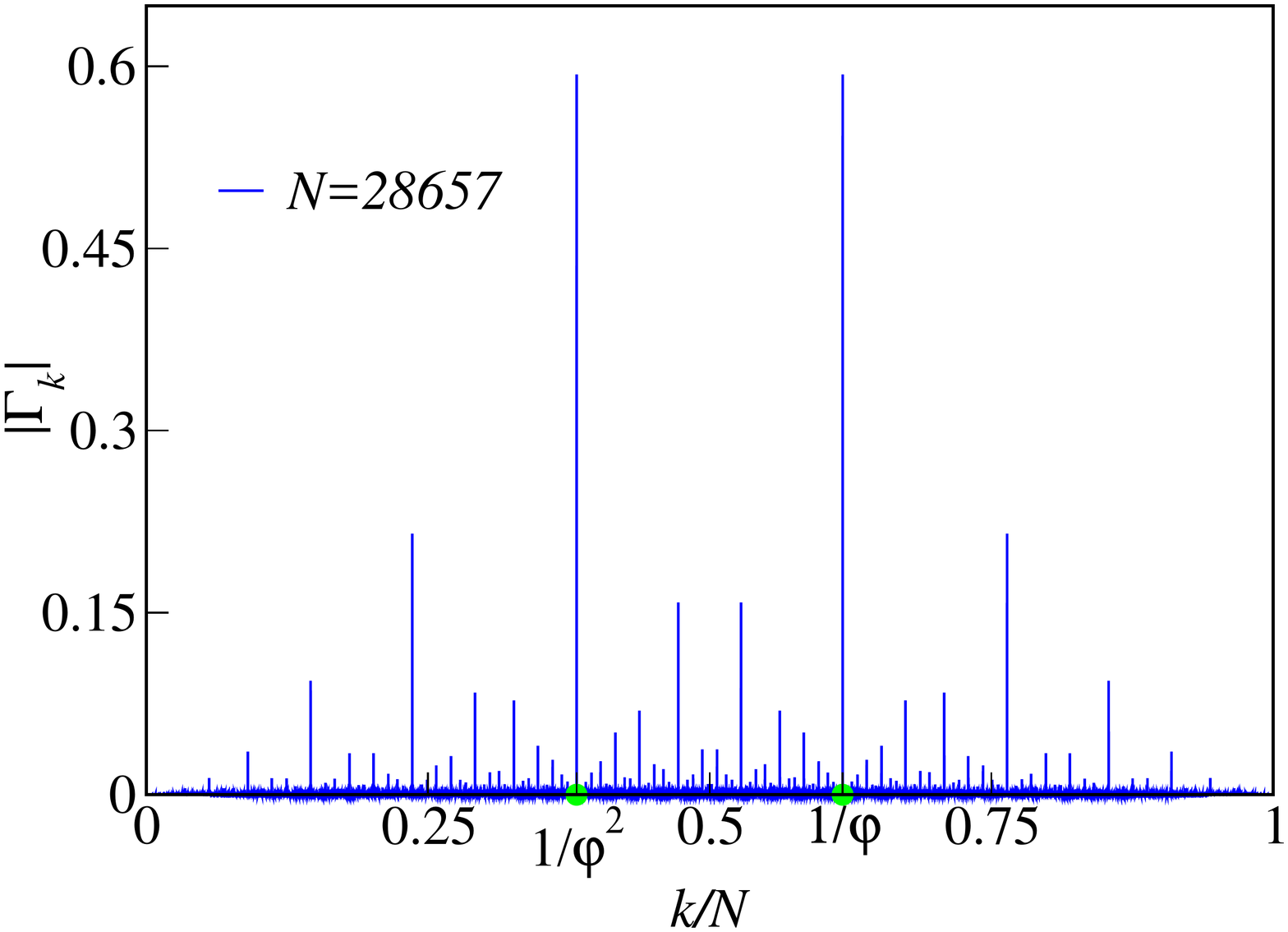}}\\ 
{\includegraphics[width=\hsize]{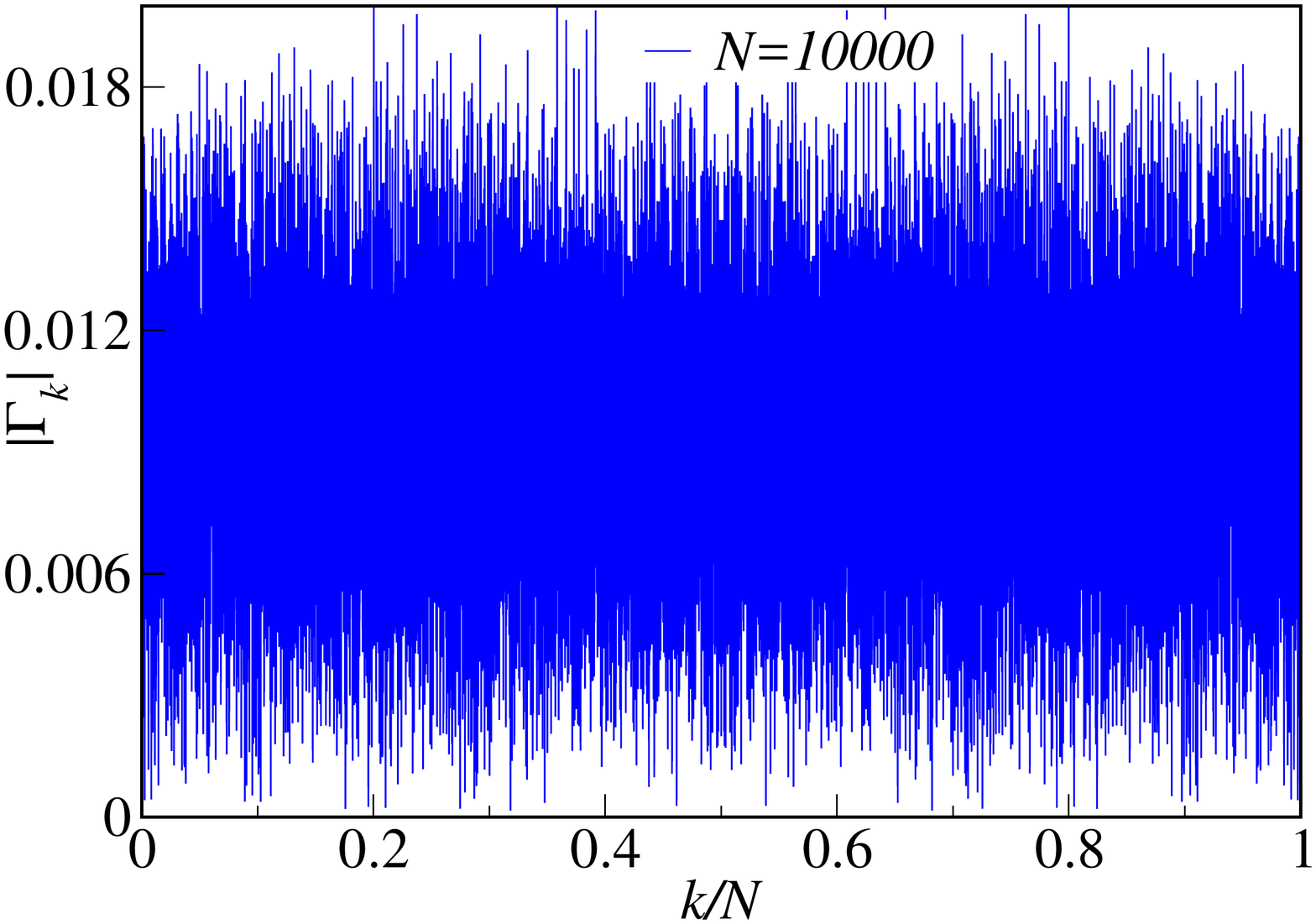}}
\caption{The magnitude of the Fourier transform $\Ga_k$ (Eq.~\eqref{FT}) 
calculated using the first $N$ terms of the sequence $\{ \tau_i \}$ as a 
function of $k/N$ where (top panel) the sequence is taken to be the Fibonacci 
sequence, and (bottom panel) the sequence is taken to be a typical realization
of a random sequence. The number of terms $N$ taken to calculate the Fourier 
transforms is indicated inside the figure panels.} \label{fig1} \end{figure}

The difference between a random sequence of $\{\tau_j\} = \pm 1$ and the 
Fibonacci sequence can be easily seen by calculating the Fourier 
transform of the two sequences defined as
\bea \Ga_k ~=~ \frac{1}{N} ~\sum_{j=1}^{N} ~\tau_{j} ~\exp \left(\frac{i2\pi k 
(j-1)}{N} \right), \label{FT} \eea
where the Fourier transform is calculated using the first $N$ terms of the 
sequence $\{ \tau_j \}$, and $k=0,1,2,\cdots,N-1$. In Fig.~\ref{fig1} we show 
the magnitude $|\Ga_k|$ as a function of $k/N$ for (a) the Fibonacci sequence 
and (b) a typical realization of a random sequence. While the Fourier transform
is featureless for the random case, the Fibonacci sequence shows sharp peaks 
at frequencies that are incommensurate with each other, e.g., the ratio of the 
frequencies of the two highest peaks equals the golden ratio $\varphi$.

\subsection{$P_\infty(\cos\theta_{k_c},\phi_{k_c})$ for p-GGE and ITE}
\label{sec:IIe}

The form of $P_\infty(\cos\theta_{k_c},\phi_{k_c})$ was discussed in 
Ref.~\onlinecite{NandySS2017} and the functions have a simple geometric 
interpretation both for a p-GGE and for an ITE which we will briefly 
recap here. 

\begin{figure}
{\includegraphics[width=\hsize]{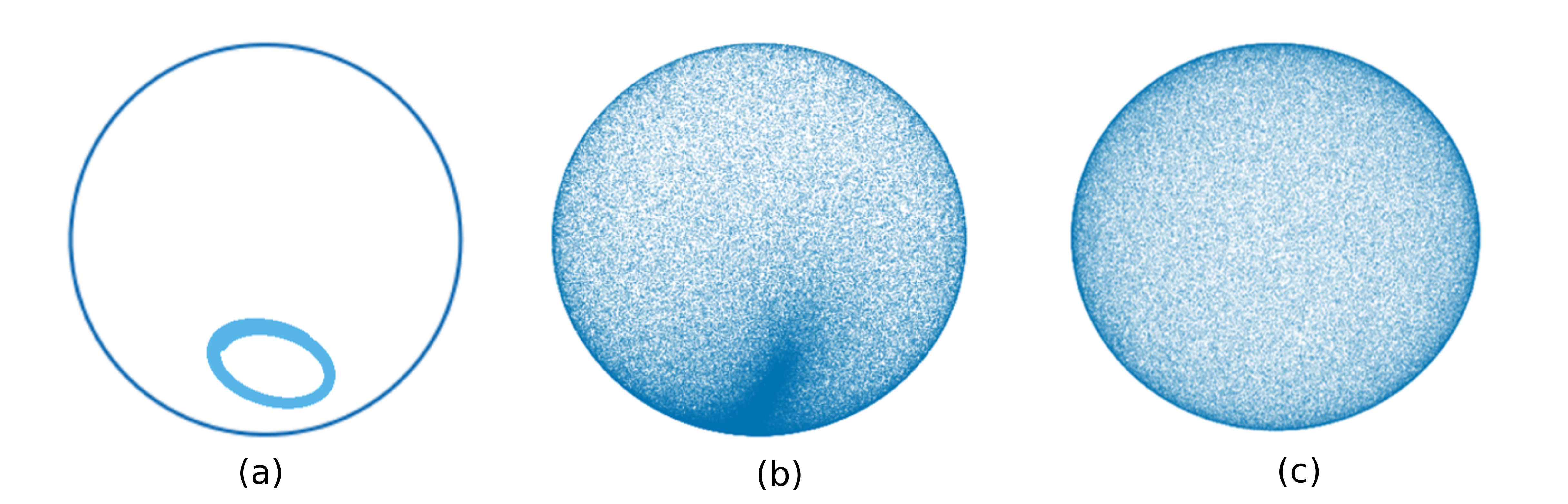}} 
\caption{Schematic distribution of the $N_c \gg 1$ points in a momentum cell 
at late times traces out a circle on the surface of the unit sphere for a 
p-GGE (shown in (a)) and covers the surface of the unit sphere uniformly for an
ITE (shown in (c)). For the quasiperiodic Fibonacci drive protocol, the $N_c$ 
points form a stable $P_\infty(\cos \theta_{k_c}, \phi_{k_c})$ that is 
intermediate between (a) and (c), as shown schematically in (b).} 
\label{fig2} \end{figure}

For a periodic drive protocol, $U_k(T)$ can be written as
\bea U_k(T) ~=~ e^{i \ga_k \hat{e}_k \cdot \vec{\sigma}}, 
\label{Uk_standardform} \eea
where $0 \leq \ga_k \leq \pi$, and $\hat{e}_k$ is a unit vector with its 
$x$, $y$, $z$ components denoted by $e_{1k}$, $e_{2k}$, $e_{3k}$ respectively, 
and $\vec{\sigma} = (\sigma^x,\sigma^y,\sigma^z)$ denote the Pauli matrices. 
Following the trajectory of a single $k$ mode $|\psi_k(n) \rangle$ as a 
function of $n$ then generates a circle on the Bloch sphere which is defined 
by the intersection of the unit sphere with a plane that contains the south 
pole (due to the initial condition $|\psi(0) \rangle$ that we have taken 
here) and whose 
normal vector equals $\hat{e}_{k}$. Taking a momentum cell whose width $\de k$ 
is small enough such that the variation in $\hat{e}_{k}$ is negligible for the 
$k$ modes inside the cell, it can then be shown~\cite{NandySS2017} that the 
$N_c \gg 1$ points (that belong to the consecutive $k$ modes of the 
coarse-grained cell) on the unit sphere uniformly cover the circle formed by 
the intersection of this unit sphere with a plane that contains the south pole 
and whose normal vector equals $\hat{e}_{k_c}$. Such a distribution is shown 
schematically in Fig.~\ref{fig2}(a) and arises since $\langle \psi_k(n)| 
\hat{e}_k \cdot \vec{\sigma} |\psi_k(n) \rangle$ is conserved at each $k$. 
For a different momentum cell with another average 
momentum $k_c$, the circle on the unit sphere changes due to the change 
in $\hat{e}_{k_c}$ as a function of $k_c$. For an ITE, the distribution of the 
$N_c$ points on the unit sphere is even simpler in that these points 
uniformly cover the surface of the unit sphere independent of $k_c$ (shown 
in Fig.~\ref{fig2}(c)). For such a case, following the trajectory of 
a single $k$ mode $|\psi_k(n) \rangle$ as a function of $n$ also 
covers the unit sphere uniformly when $n$ is large enough. 
We will show below that for the quasiperiodically driven case,
the $N_c$ points in a momentum cell form a stable distribution $P_\infty (\cos
\theta_{k_c}, \phi_{k_c})$ that is neither a circle (Fig.~\ref{fig2}(a)) nor a
uniform cover (Fig.~\ref{fig2}(c)) on the unit sphere surface but is 
intermediate between these two extreme cases (shown schematically in
Fig.~\ref{fig2}(b)). In fact, depending on the exact drive parameters 
(including the form of the drive protocol) and and the value of $k_c$, 
a Fibonacci drive 
protocol may actually give a wide variety of $P_\infty (\cos \theta_{k_c}, 
\phi_{k_c})$ ranging all the way from Fig.~\ref{fig2}(a) to Fig.~\ref{fig2}(c).
From this geometric viewpoint, it is clear how such a NESS formed due to this 
quasiperiodic drive protocol is qualitatively different from either a p-GGE or 
an ITE and allows for much richer possibilities as a function of the drive 
parameters as we show in the rest of the paper.

\section{Results for quasiperiodically driven 1D TFIM}
\label{sec:III}

\subsection{Behaviour of quantities as a function of $n$}
\label{sec:IIIa}

We now study the unitary dynamics of the 1D TFIM when $dT \neq 0$ and the 
$\tau_i$'s in Eq.~\eqref{perturbedFloquet} are given by the Fibonacci sequence 
described in Eq.~\eqref{recursiveF}. 
\begin{figure}
{\includegraphics[width=\hsize]{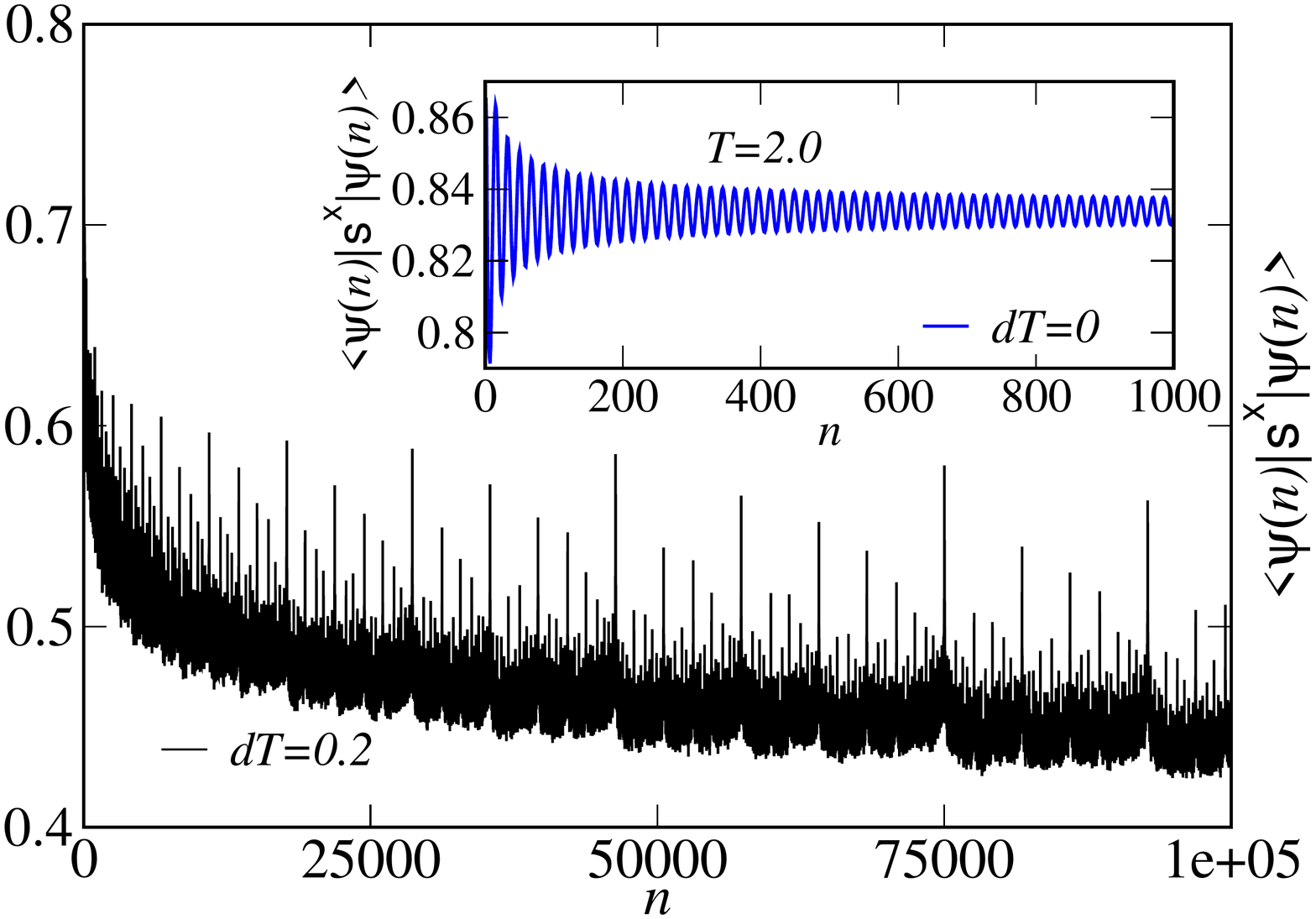}} \\
{\includegraphics[width=\hsize]{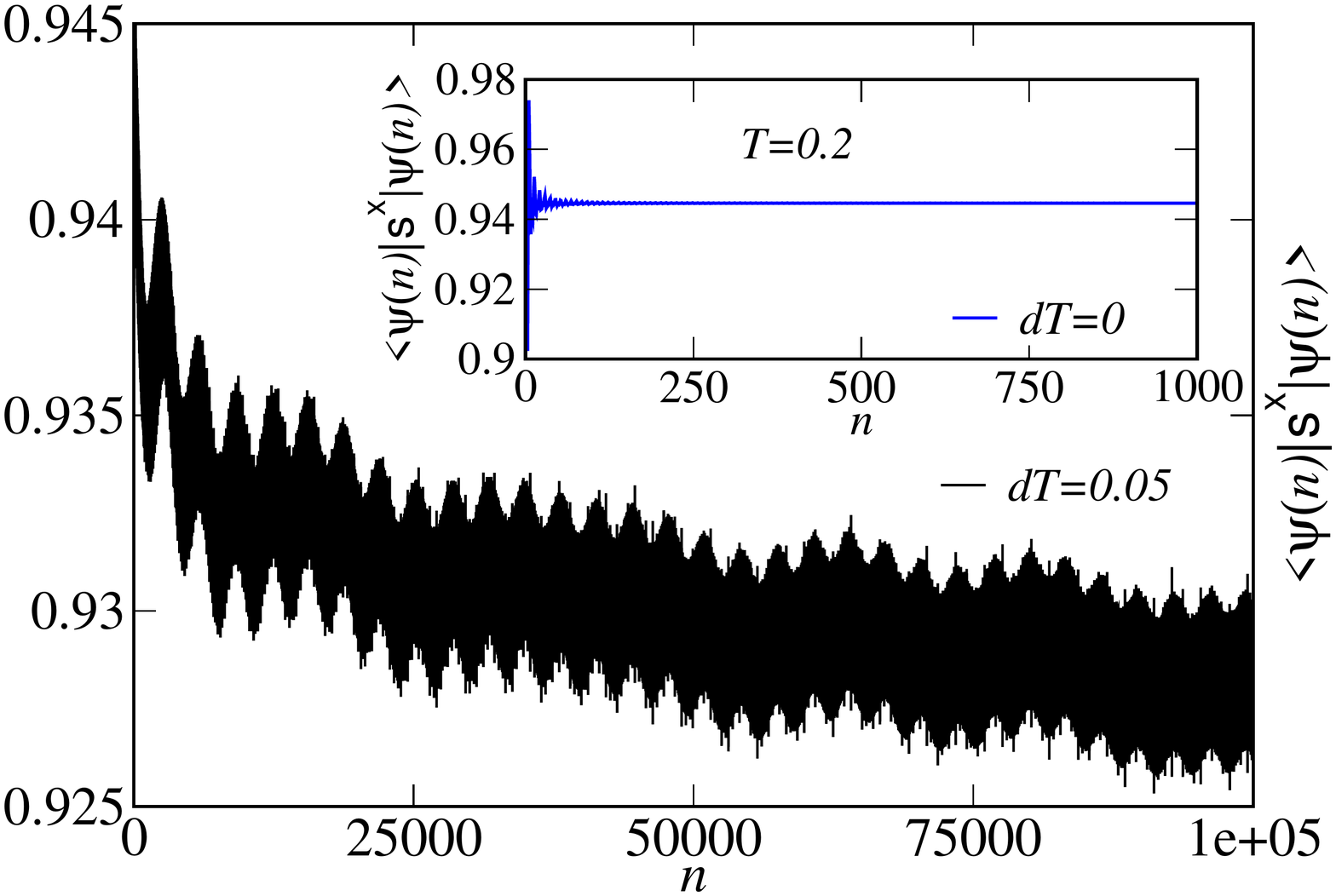}}
\caption{The behaviour of $\langle \psi(n)|s^x|\psi(n) \rangle$ as a function
of $n$ for the Fibonacci quasiperiodic drive protocol with the parameters 
$g_i=4$, $g_f=2$ and $T=2.0$, $dT=0.2$ (top panel), $T=0.2$, $dT=0.05$ 
(bottom panel). The system size equals $L=4194304$. The inset in both panels
shows the behaviour of $\langle \psi(n)|s^x|\psi(n) \rangle$ for the perfectly 
periodic drive with the corresponding value of $T$.} \label{fig3} \end{figure}

We first study a local quantity, $\langle \psi(n)|s^x|\psi(n) \rangle$, as a 
function of the drive number $n$. This can be straightforwardly 
calculated in the fermionic representation (Eq.~\eqref{jw1}). 
We show the results for two cases in Fig.~\ref{fig3} where the drive
parameters are $g_i=4$, $g_f=2$, with $T=2.0$ and $dT=0.2$ in the top panel 
and $T=0.2$, $dT=0.05$ in the bottom panel. The behaviour of this quantity for 
the same $T$ but with $dT=0$, i.e., a perfectly periodic drive protocol, 
is also shown in the inset of both panels of Fig.~\ref{fig3}. Unlike the 
periodically driven system where $\langle \psi(n)|s^x|\psi(n) \rangle$ 
approaches a steady state value (described by the corresponding p-GGE and 
shown in the insets of the top and bottom panels), this local
quantity does not seem to approach any well-defined steady state value as a
function of $n$ for the quasiperiodically driven system even for $n \sim 10^5$
(Fig.~\ref{fig3}). Furthermore, $\langle \psi(n)|s^x|\psi(n) \rangle$ has 
cusp-like singularities at multiple values of $n$ (clearly visible in 
Fig.~\ref{fig3} (top panel) but also present in Fig.~\ref{fig3} (bottom panel)) 
with the strongest features being present when $n$ equals a Fibonacci number. 

\begin{figure}
{\includegraphics[width=\hsize]{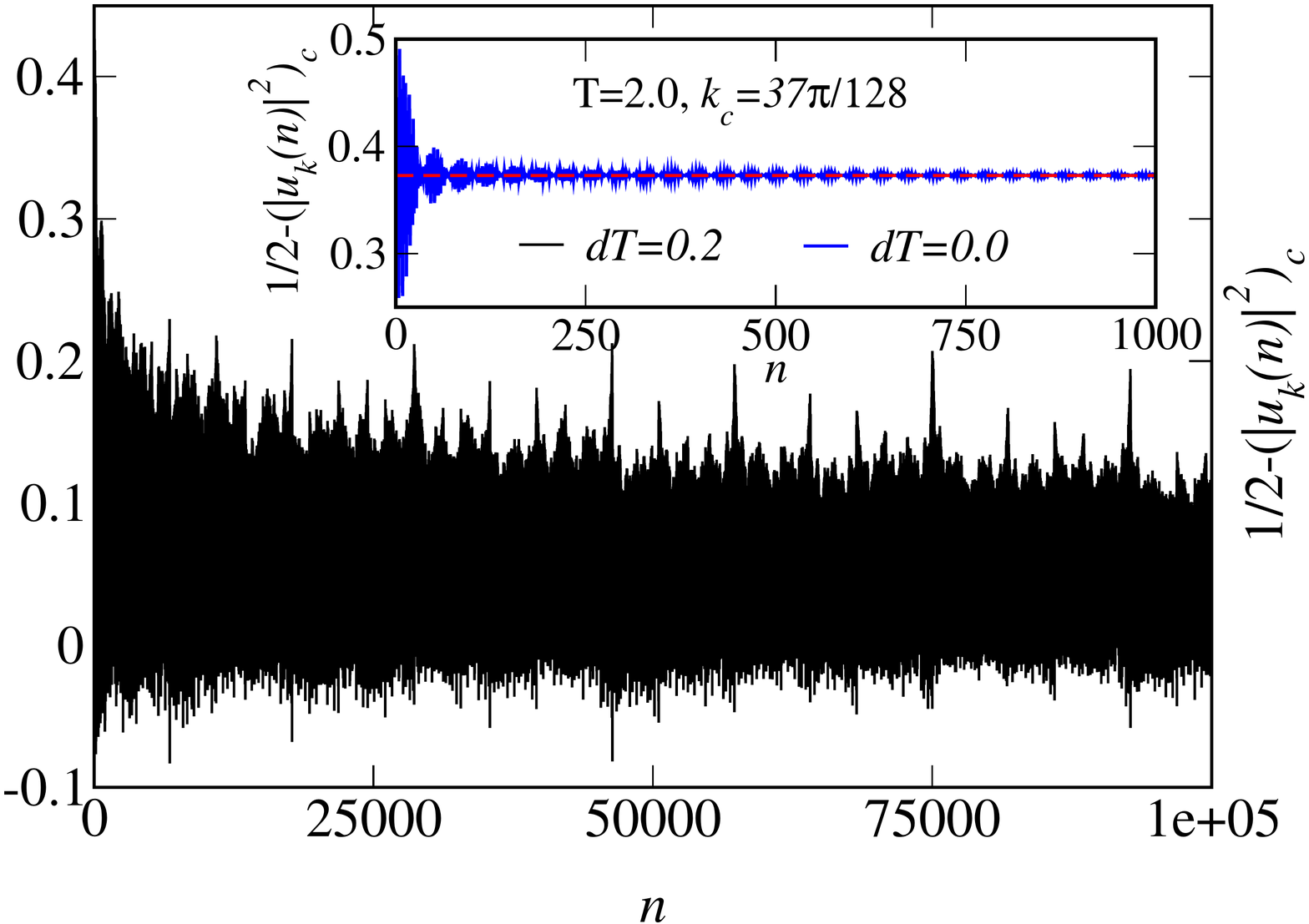}} \\
{\includegraphics[width=\hsize]{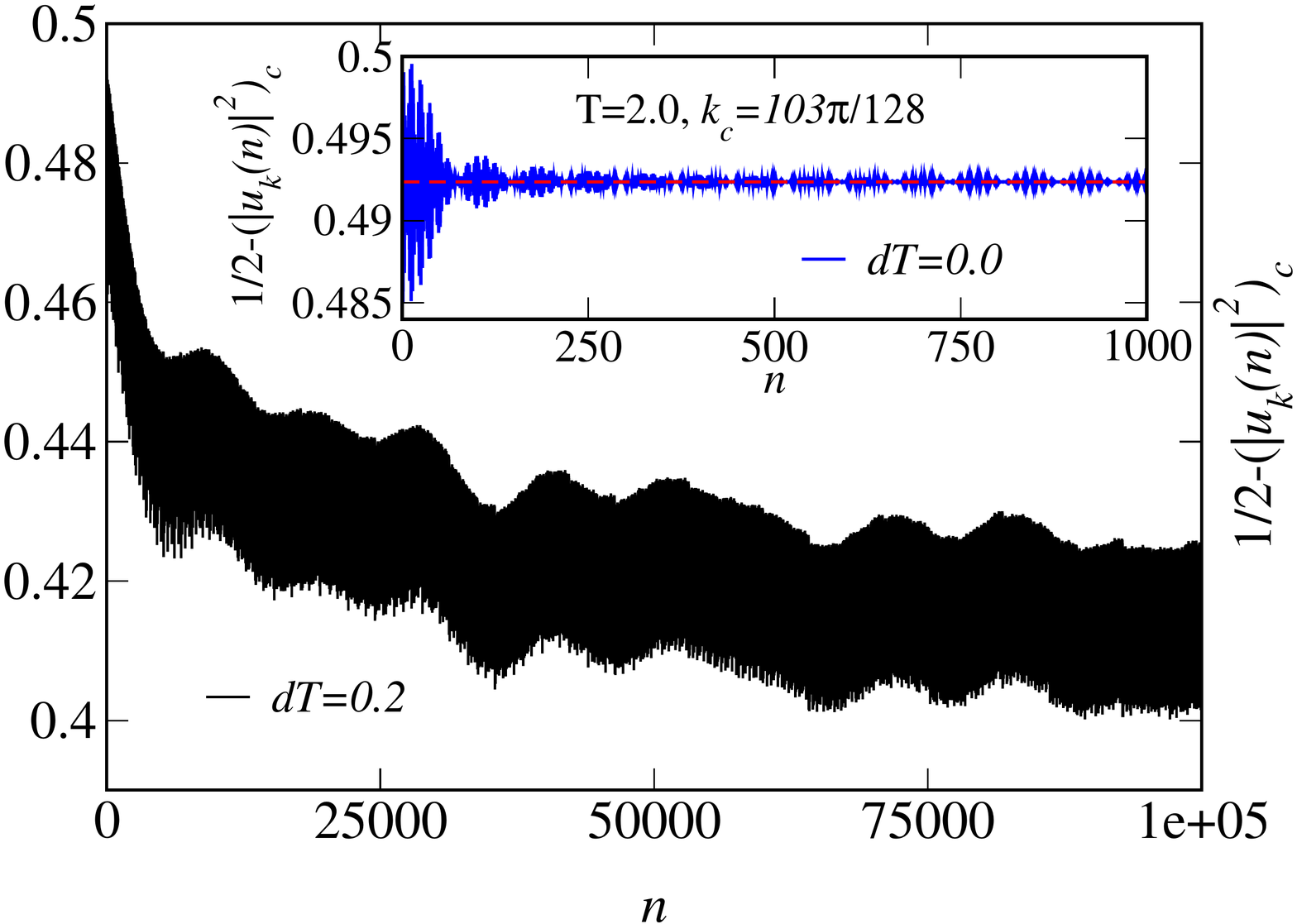}}
\caption{The behaviour of $\left(|u_k(n)|^2\right)_c$ as a function on $n$ for 
the Fibonacci quasiperiodic drive protocol with the parameters $g_i=4$, $g_f=2$,
$T=2.0$ and $dT=0.2$. The system size equals $L=4194304$ and the $k$ space
is divided into $64$ cells with $32768$ consecutive momenta each. The top
panel displays the result for $k_c=37\pi/128$ and the bottom panel for
$k_c=103\pi/128$, where $k_c$ denotes the average momentum of the
corresponding momentum cell. The inset of both panels shows the behaviour of 
the corresponding quantity for a periodically driven system with $dT=0$. 
The horizontal red line in both insets indicate the expected steady state 
value for $dT=0$.} \label{fig4} \end{figure}

We also show the behaviour of the coarse-grained quantities $(|u_k(n)|^2)_c$ 
defined in Eq.~\eqref{coarsegrain} as a function of $n$ for the drive protocol 
with $T=2.0$ and $dT=0.2$ ($g_i=4$, $g_f=2$) in Fig.~\ref{fig4}. As explained 
earlier, these quantities approach well-defined steady state values for $n
\to \infty$ if a NESS exists. For example, $1/2-(|u_k(n)|^2)_c \to 0$ for $n 
\to \infty$ for any coarse-grained momentum cell if the NESS is an ITE. 
Similarly, $1/2-(|u_k(n)|^2)_c\to \left( \frac{1}{2}\right) e^2_{3k_c}$ 
(see Eq.~\eqref{Uk_standardform}) for a periodically driven 
system~\cite{NandySS2017} provided $\de k$ is small enough so that the 
variation in $\hat{e}_k$ is negligible for the momenta inside a coarse-grained 
cell. To construct these coarse-grained quantities, we use a system size 
of $L=4194304$ spins and divide the $k$ space
into $64$ momentum cells with each cell containing $32768$ consecutive
momenta. In Fig.~\ref{fig4}, we show the behaviour of $1/2-(|u_k(n)|^2)_c$ as
a function of $n$ for $k_c=37\pi/128$ (top panel) and $k_c=103\pi/128$ (bottom
panel), where $k_c$ equals the average momentum of a coarse-grained cell in
$k$ space. We also show the behaviours of the corresponding quantities for 
$dT=0$ in the inset of both the panels which show the convergence to the 
expected values for the periodically driven case. We again see that 
$(|u_k(n)|^2)_c$ does not seem to approach any well-defined steady state value 
even for $n \sim 10^5$ for both the momentum cells. The coarse-grained 
quantities also have cusp-like features at multiple values of $n$ with the 
strongest features at values of $n$ that equal any of the Fibonacci numbers, 
exactly like the local quantity $\langle \psi(n)|s^x|\psi(n) \rangle$.

\begin{figure}
{\includegraphics[width=0.6\hsize]{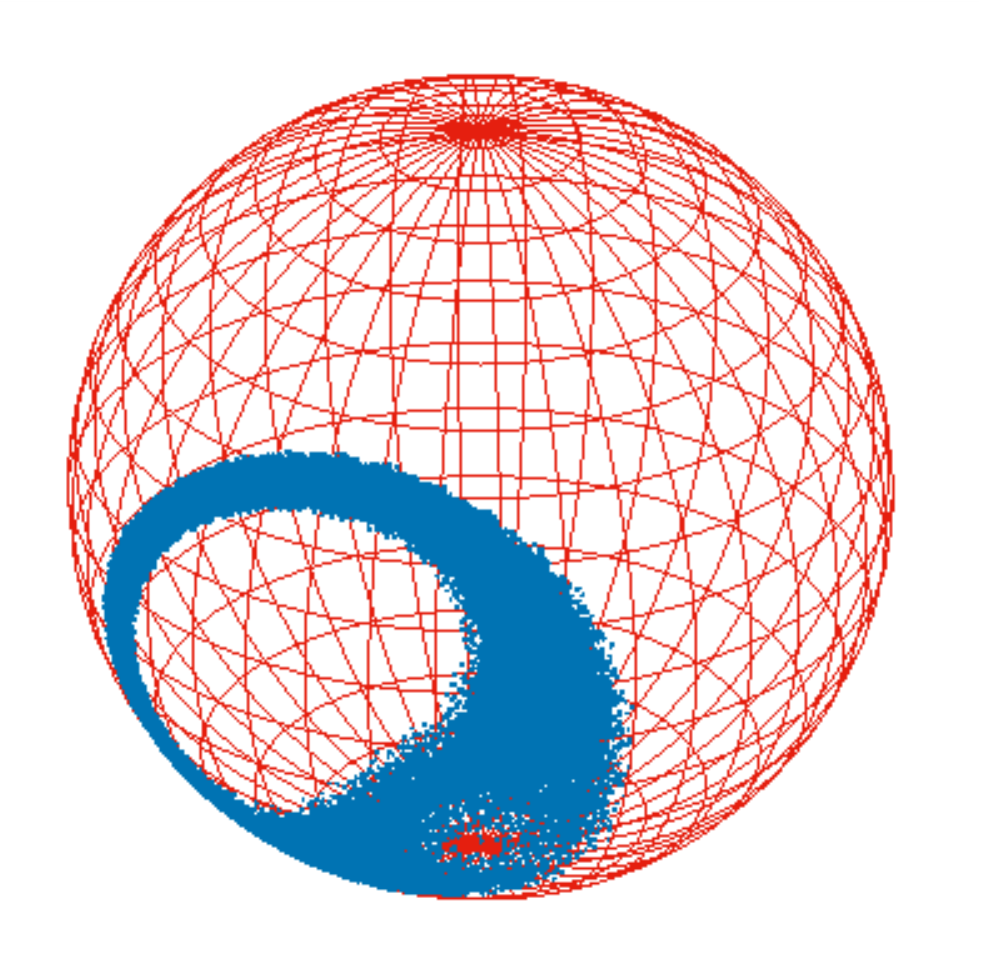}} 
\caption{The trajectory of $|\psi_k(n) \rangle$ (where $|\psi_k(n=0)\rangle = 
(0,1)^T$) for the Fibonacci drive protocol (Eq.~\eqref{perturbedFloquet}) shown 
on the Bloch sphere. The drive parameters are $g_i=4$, $g_f=2$, $T=0.2$ and 
$dT=0.05$. Here $k$ equals $\pi/2$ and the number of drives is taken to 
be $121393$.} \label{fig5} \end{figure}

Lastly, we show the trajectory of a single $k$ mode on the corresponding 
Bloch sphere as a function of $n$ in Fig.~\ref{fig5}, where we take $g_i=4$, 
$g_f=2$, $T=0.2$ and $dT=0.05$. We see that the trajectory neither follows 
a circle (as expected for a periodically driven system) nor covers the entire 
surface of the sphere uniformly (as expected for an ITE) but is a 
complicated intermediate structure.

\subsection{Behaviour of quantities as a function of $n=F_m$}
\label{sec:IIIb}

\begin{figure}
{\includegraphics[width=\hsize]{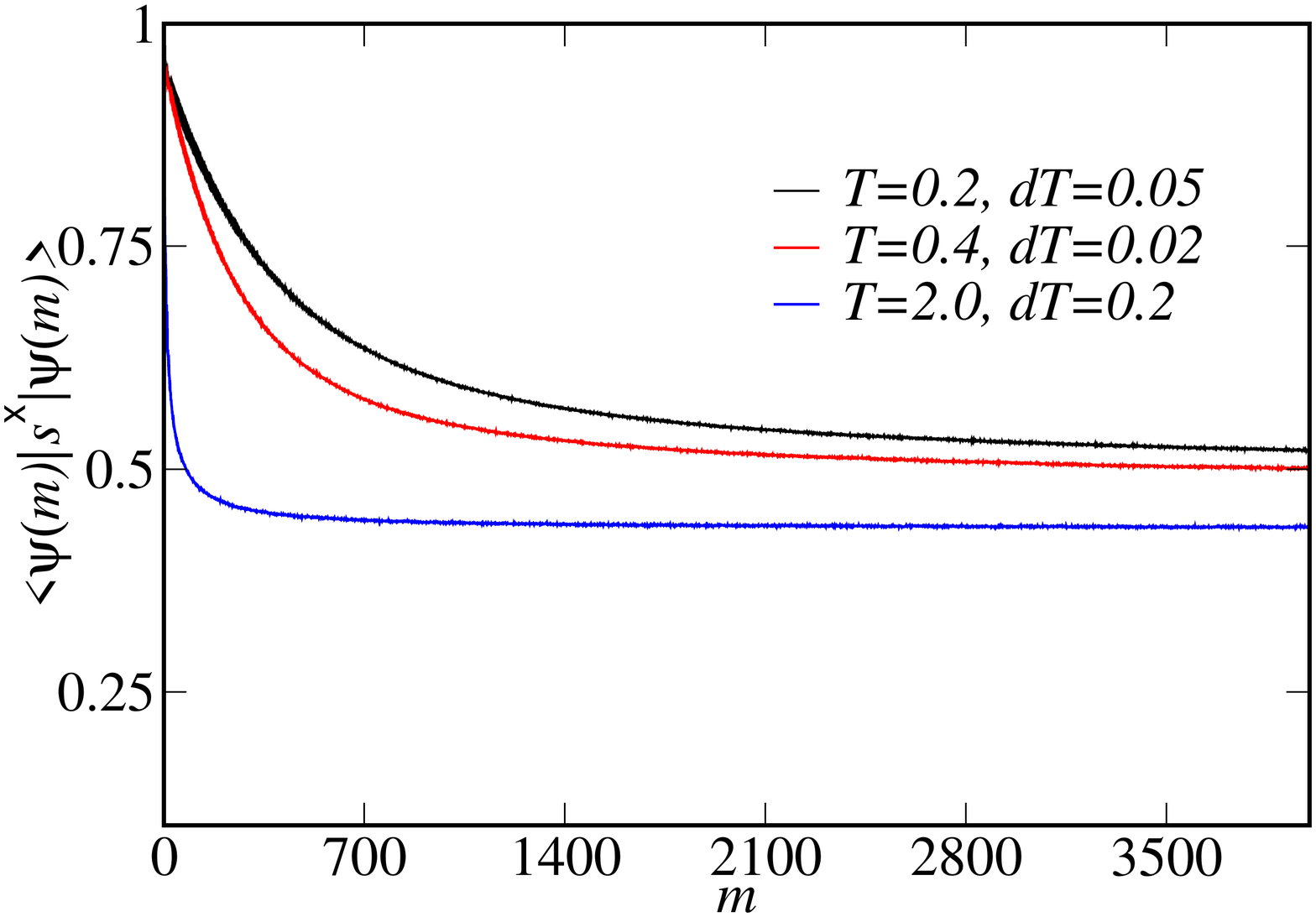}} \\
{\includegraphics[width=\hsize]{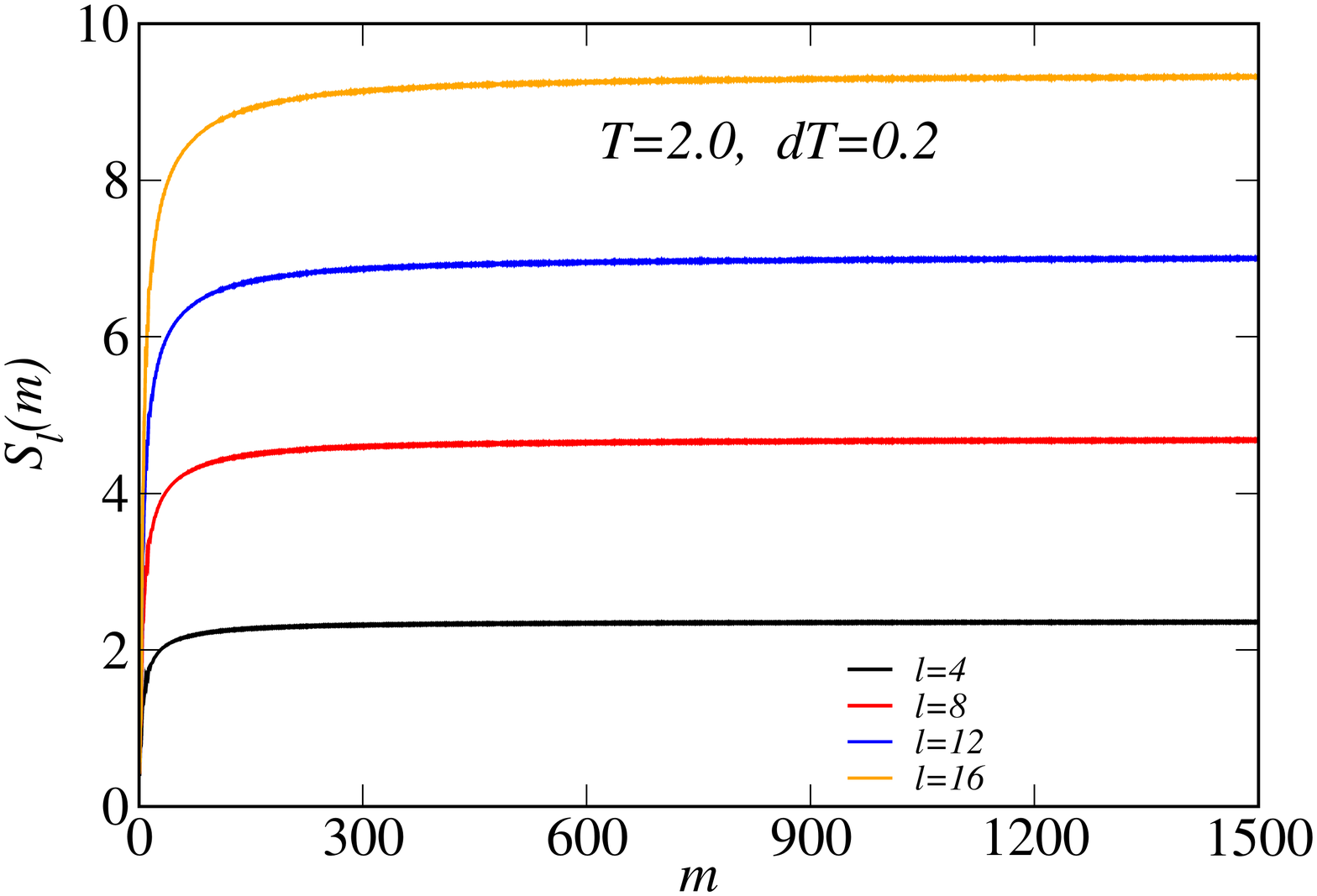}}
\caption{(Top panel) The behaviour of $\langle \psi(n)|s^x|\psi(n)\rangle$ 
as a function of $m$ where the drive number $n=F_m$ 
(we note that $n \sim \varphi^m$ for large $m$).
(Bottom panel) The behaviour of the entanglement entropy for subsystem sizes
$l=4,8,12,16$ as a function of $m$. The system size is $L=4194304$ and 
$g_i=4$ and $g_f=2$ for all the cases shown.} \label{fig6} \end{figure}

From the behaviours of the local quantities shown in Fig.~\ref{fig3} and 
Fig.~\ref{fig4}, it is clear that these do not approach steady state values 
even for $n \sim 10^5$. This is reminiscent of the case studied in 
Ref.~\onlinecite{NandySS2017} where the periodically driven system 
is perturbed with 
a noise that is scale-invariant in time. The NESS is then only approached after 
an exponentially long time when the system is viewed not stroboscopically but 
as $m=2^n$ (this was dubbed as a geometric generalized Gibbs ensemble in 
Ref.~\onlinecite{NandySS2017}). 

We now use the recursive structure of the Fibonacci sequence to study the
unitary dynamics of the system for exponentially long times $n=F_m \sim 
\varphi^m$ using only $\mathcal{O}(m)$ unitary matrix multiplications at each 
$k$. To see this, we simply note that the unitary evolution matrix at
level $m$, which we denote by $U_m$, equals 
\bea U_{m+2} ~=~ U_{m}U_{m+1}, \label{Fibonacci_U} \eea 
for all $m \geq 1$ and $U_1=U_k(T+dT)$, $U_2=U_k(T-dT)U_k(T+dT)$ (see 
Eq.~\eqref{recursiveF}). We use the recursion level index $m$ 
(Eq.~\eqref{recursiveF}) as a shorthand to denote $n = F_m$. We show results 
for $\langle \psi(n)|s^x|\psi(n) \rangle$ as 
a function of $m$ for various drive 
parameters $(T,dT)$ at $g_i=4$, $g_f=2$ in Fig.~\ref{fig6} (top panel). From 
these results, it is clear that this local operator {\it does} reach a steady 
state when the system is observed not stroboscopically but at drive numbers 
$n=F_m$. In Fig.~\ref{fig6} (top panel), we see that for some drive 
parameters ($T=2.0$, $dT=0.2$), the steady state is reached much sooner with
$m$ than for the other drive parameters ($T=0.2$, $dT=0.05$ and $T=0.4$, 
$dT=0.02$). Thus, the value of $\langle \psi(n)|s^x|\psi(n) \rangle$ shown in 
Fig.~\ref{fig3} (bottom panel) for $T=0.2$, $dT=0.05$ up to $n \sim 10^5$ is 
nowhere close to the final steady steady value. While these results strongly 
suggest a NESS for $l=1$ subsystems, it is also essential to establish the 
same for other subsystem sizes (where $l \ll L$). To this end, we calculate 
the entanglement entropy $S_{\mathrm ent}(l)$ (Eq.~\eqref{entanglement}) for 
subsystems of size $l=4,8,12,16$ as a function of $n=F_m$ and show the results 
in Fig.~\ref{fig6} (bottom panel) for $T=2.0$ and $dT=0.2$. The entanglement 
entropy for all the subsystem sizes saturate to well-defined steady state 
values as a function of $n=F_m$ which shows that a NESS (as far as local 
quantities are concerned) is indeed reached
at exponentially long times. We also note that $S_{\mathrm ent}(l) \sim l$ for 
the steady state values of the entanglement entropy in Fig.~\ref{fig6} (bottom 
panel) and thus follows a volume law scaling expected for a NESS. 

\begin{figure}
{\includegraphics[width=\hsize]{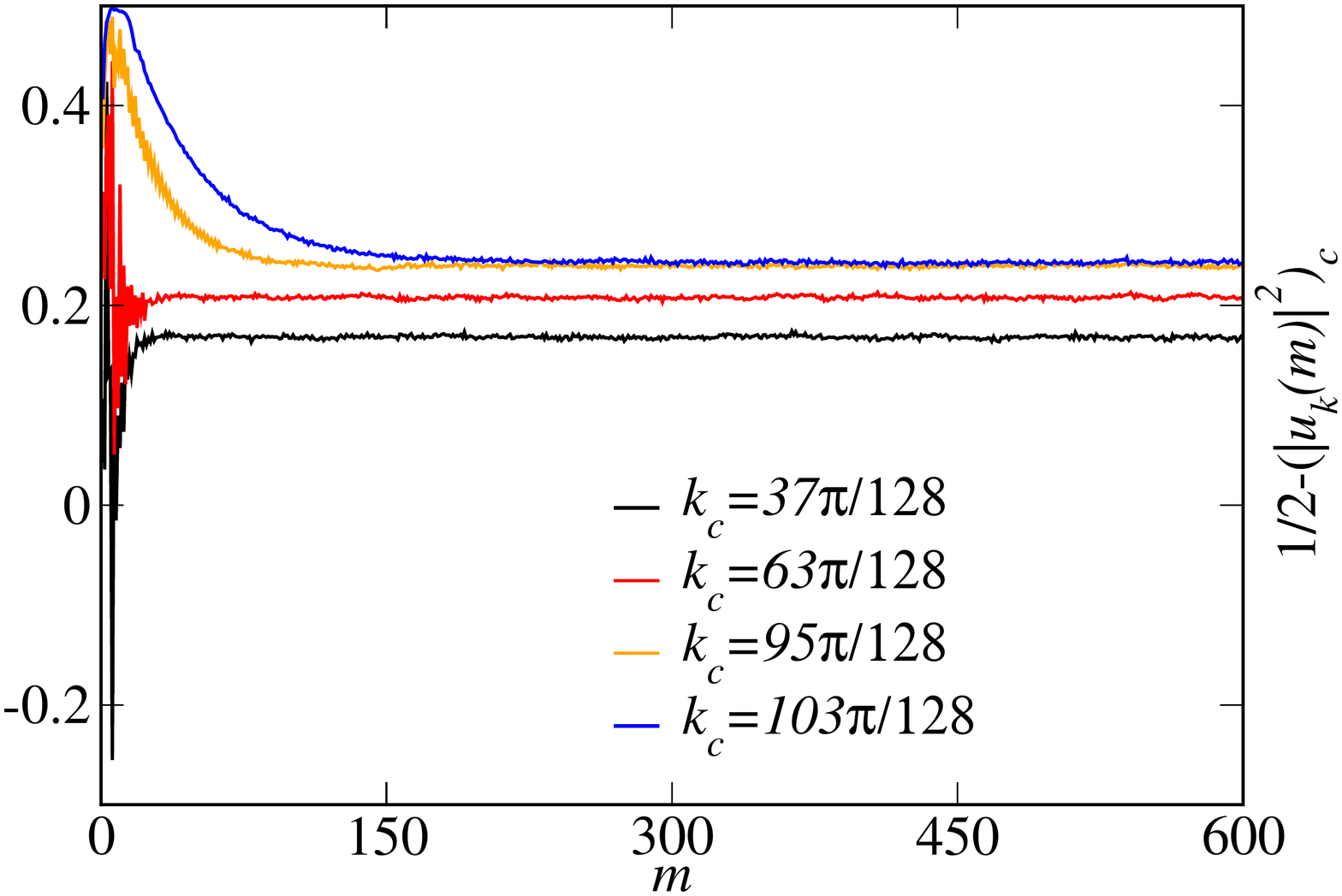}}
\caption{The behaviour of the coarse-grained quantities $1/2-(|u_k(n)|^2)_c$ 
as a function of $m$ where the drive number $n=F_m$.
The system size is $L=4194304$ and the drive parameters $g_i=4$, $g_f=2$, 
$T=2.0$ and $dT=0.2$. The momentum space is divided into $64$ cells with each 
cell having $32768$ consecutive momentum modes to construct these 
coarse-grained quantities. $k_c$ denotes the average momentum of a 
coarse-grained cell.} \label{fig7} \end{figure}

Further evidence for the emergence of a NESS is obtained by looking at the 
behaviour of the coarse-grained quantities $1/2-(|u_k(n)|^2)_c$ for different 
momentum cells as a function of $n=F_m$ instead of stroboscopically 
(Fig.~\ref{fig7}) for the drive parameters $g_i=4$, $g_f=2$, $T=2.0$ and 
$dT=0.2$. These coarse-grained quantities also reach a steady state 
as a function of $n=F_m$ with the steady state value being a function of 
$k_c$. Furthermore, the convergence of $(|u_k(n)|^2)_c$ to its steady 
state value is a strong function of $k_c$ (Fig.~\ref{fig7}) which implies 
that different local operators have different time scales of 
approach to their corresponding steady state values. 

\subsection{Behaviour of $\mathcal{P}_\infty(\cos \theta_{k_c},\phi_{k_c})$}
\label{sec:IIIc}

\begin{figure}
{\includegraphics[width=0.49 \columnwidth]{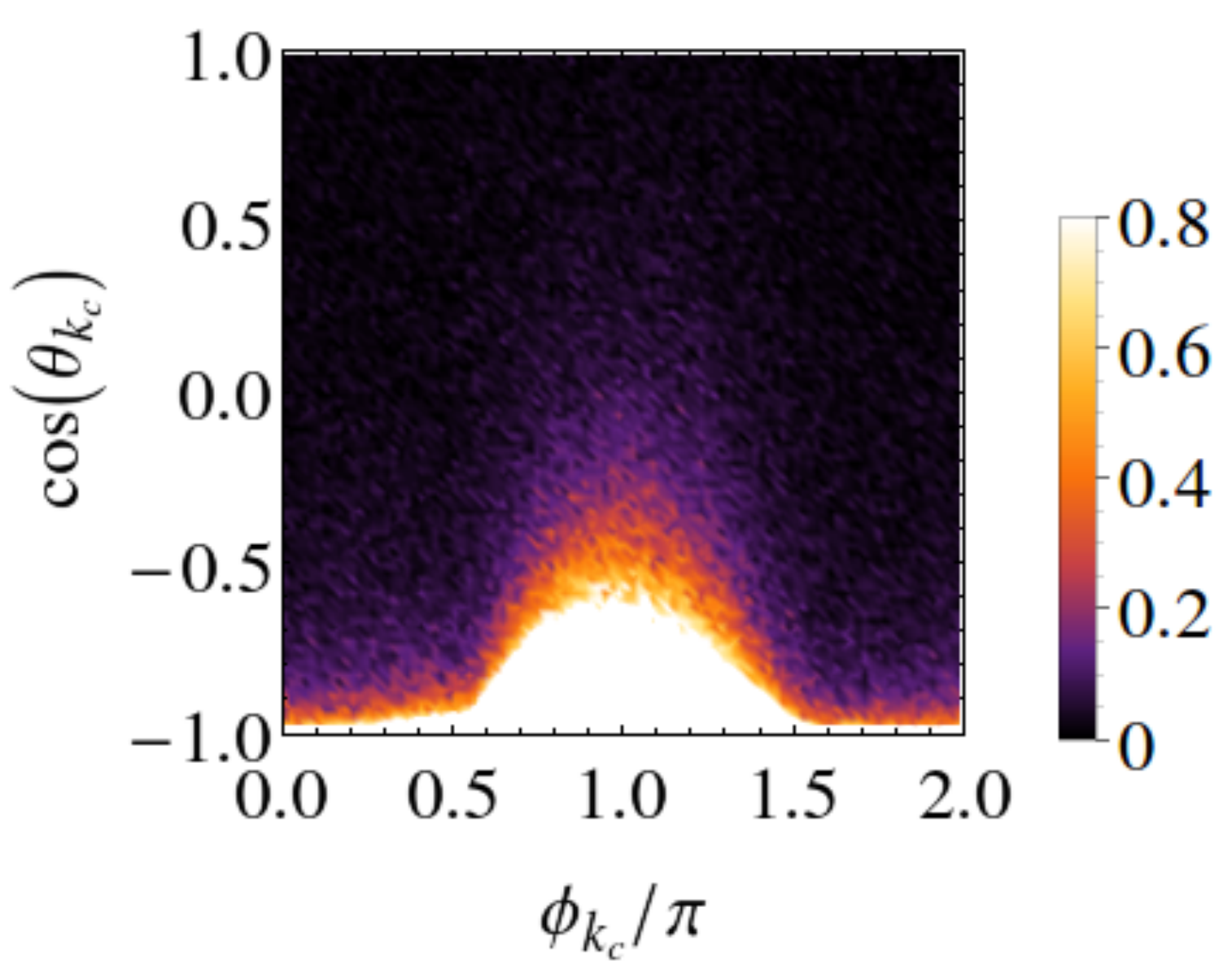}}%
{\includegraphics[width=0.49 \columnwidth]{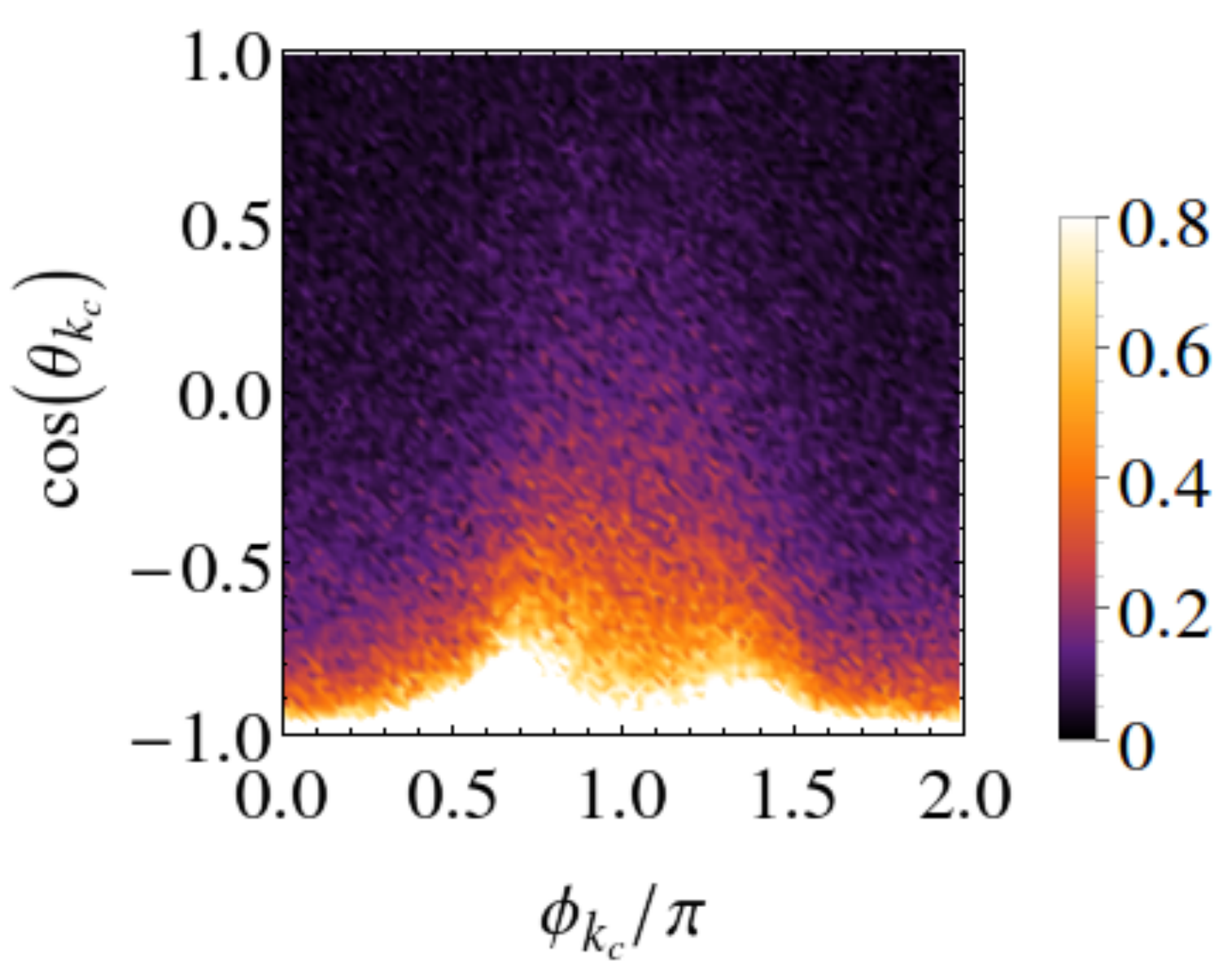}}\\
{\includegraphics[width=0.49 \columnwidth]{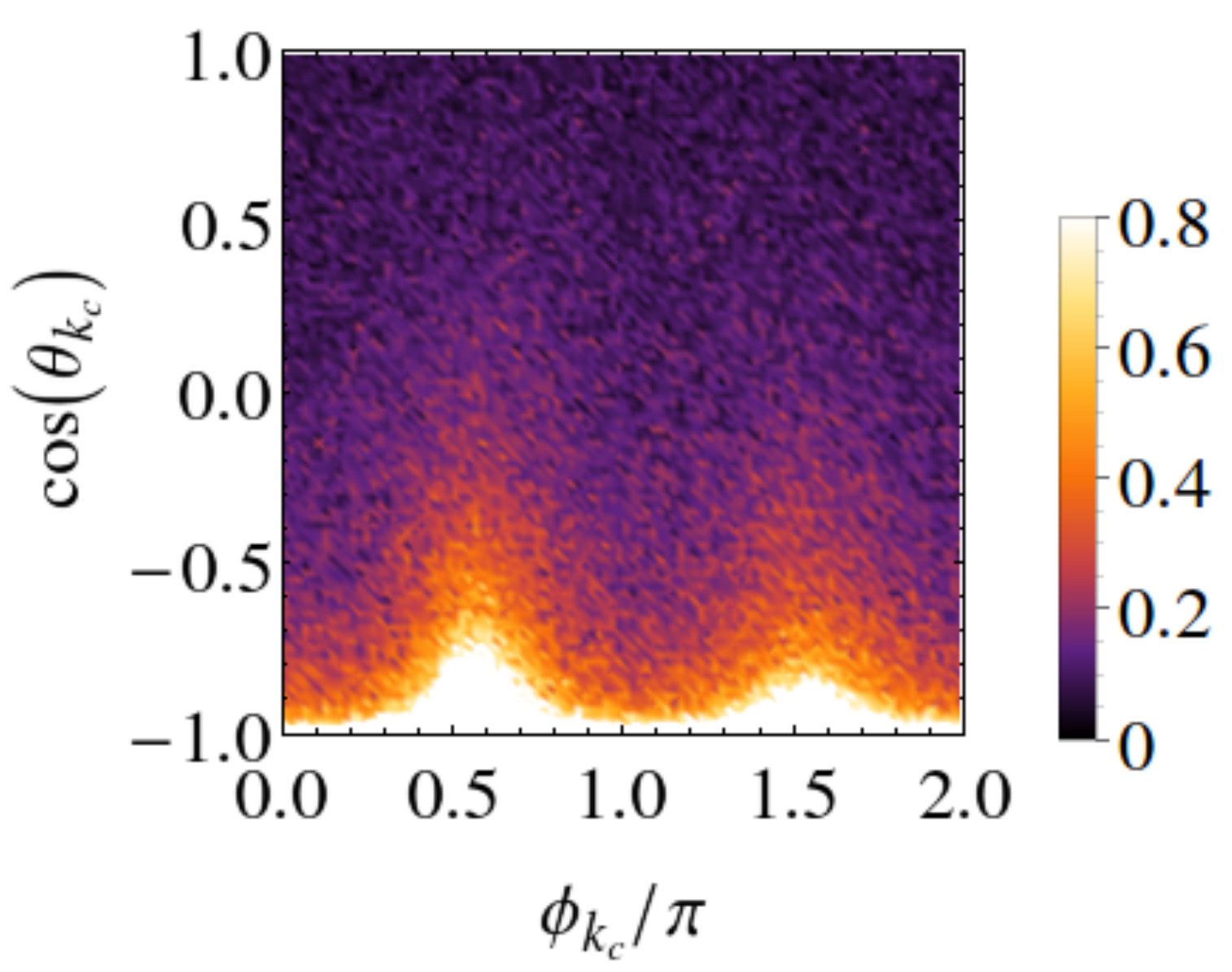}}%
{\includegraphics[width=0.49 \columnwidth]{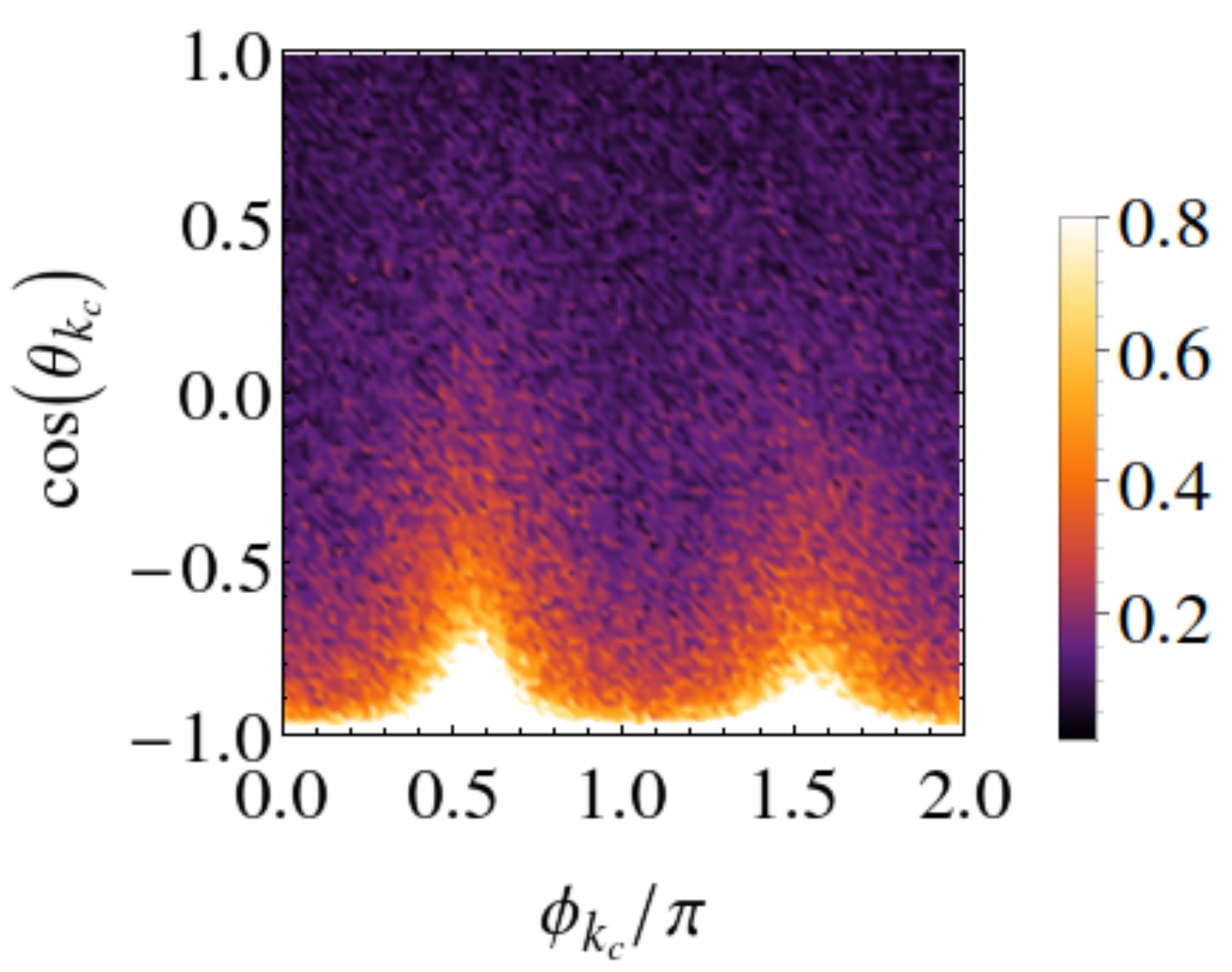}} 
\caption{The evolution of $\mathcal{P}_n(\cos \theta_{k_c},\phi_{k_c})$ as a 
function of $n=F_m$ for a coarse-grained momentum cell with $N_c=262144$ 
consecutive momenta and $k_c=31\pi/64$. The drive parameters are $g_i=4$, 
$g_f=2$, $T=0.2$ and $T=0.05$. 
$m$ takes the value $100$ (top left), $300$ (top right), $1000$ (bottom 
left), $2000$ (bottom right) in the four panels.} \label{fig8} \end{figure}

\begin{figure}
{\includegraphics[width=0.49 \columnwidth]{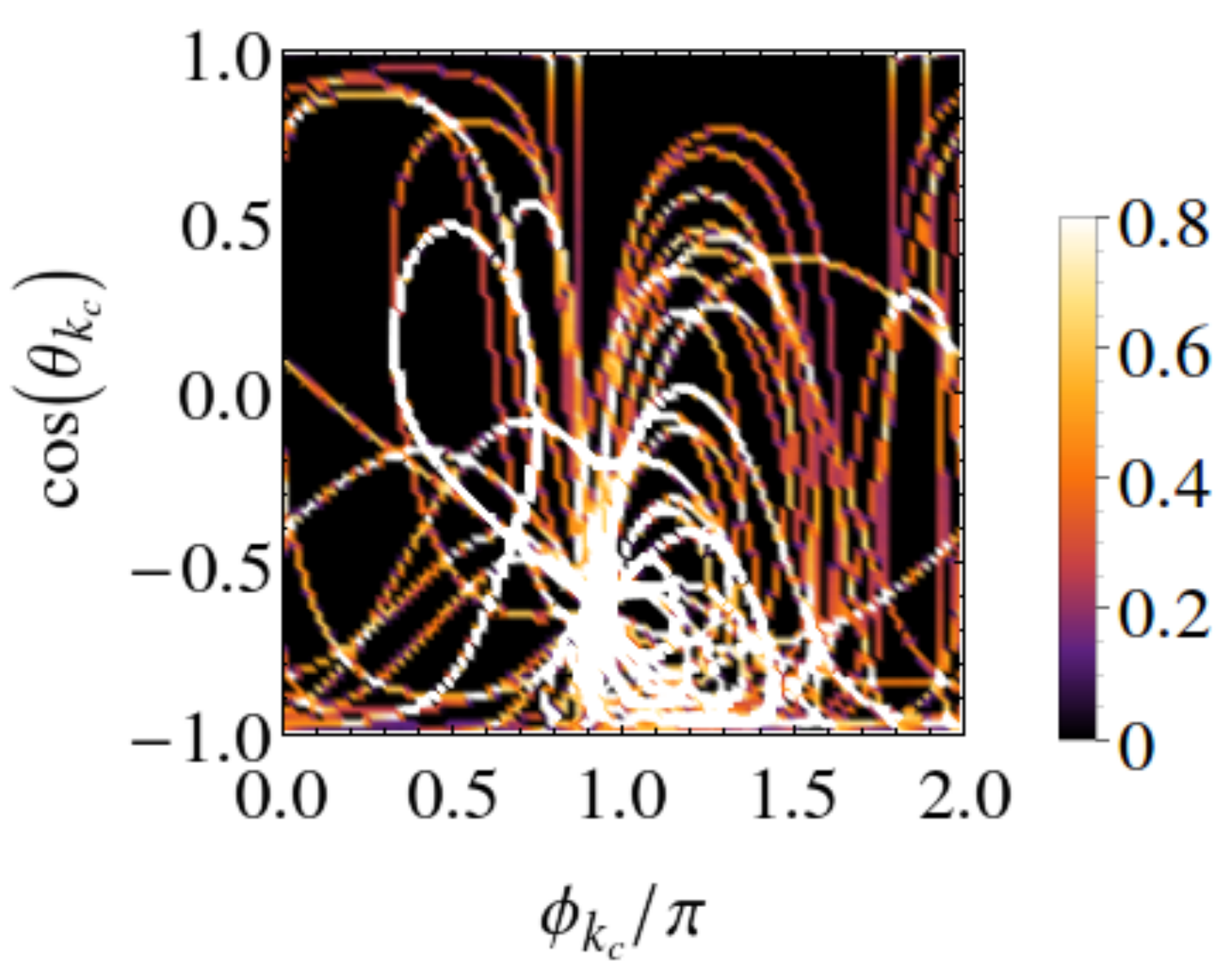}}%
{\includegraphics[width=0.49 \columnwidth]{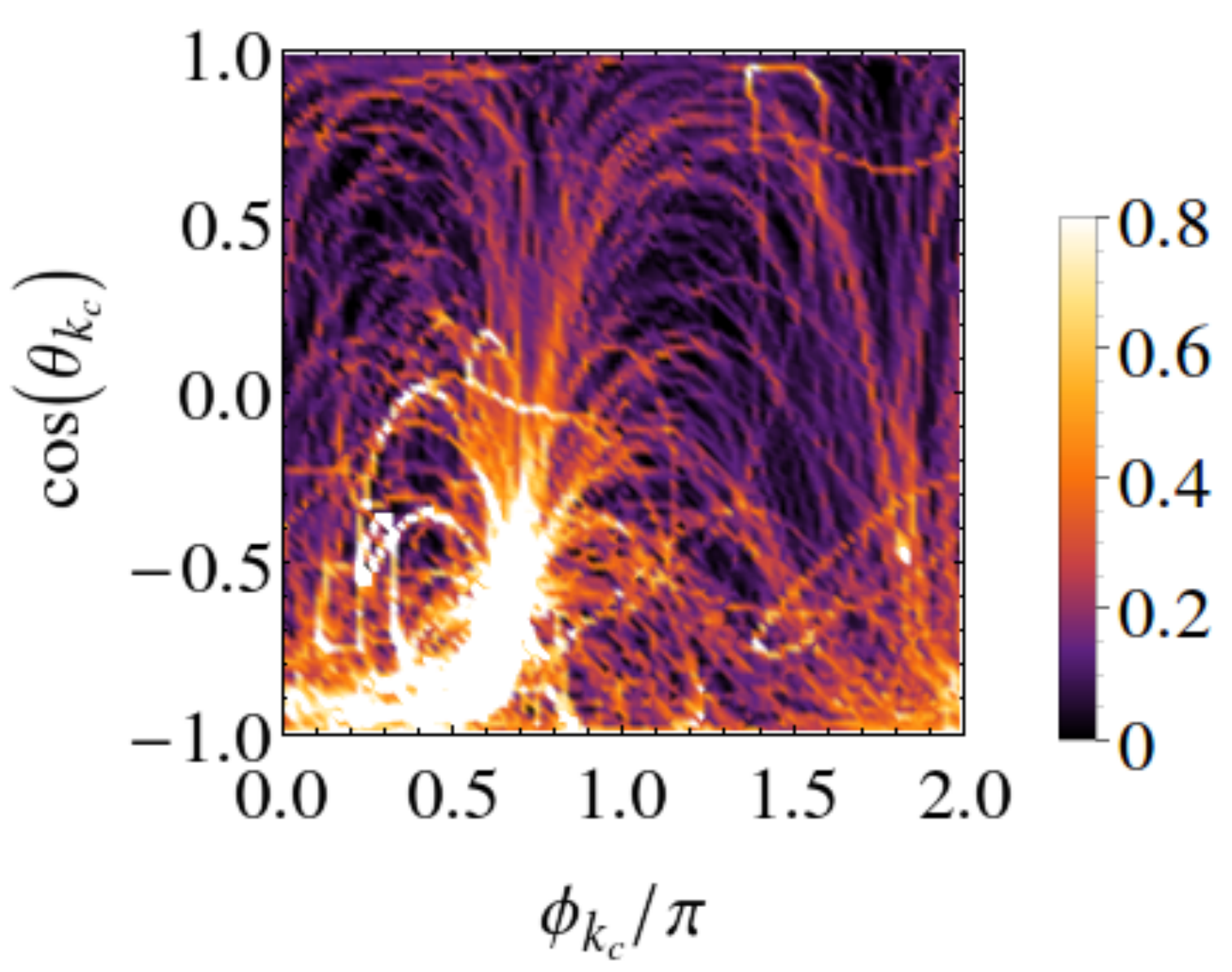}}\\
{\includegraphics[width=0.49 \columnwidth]{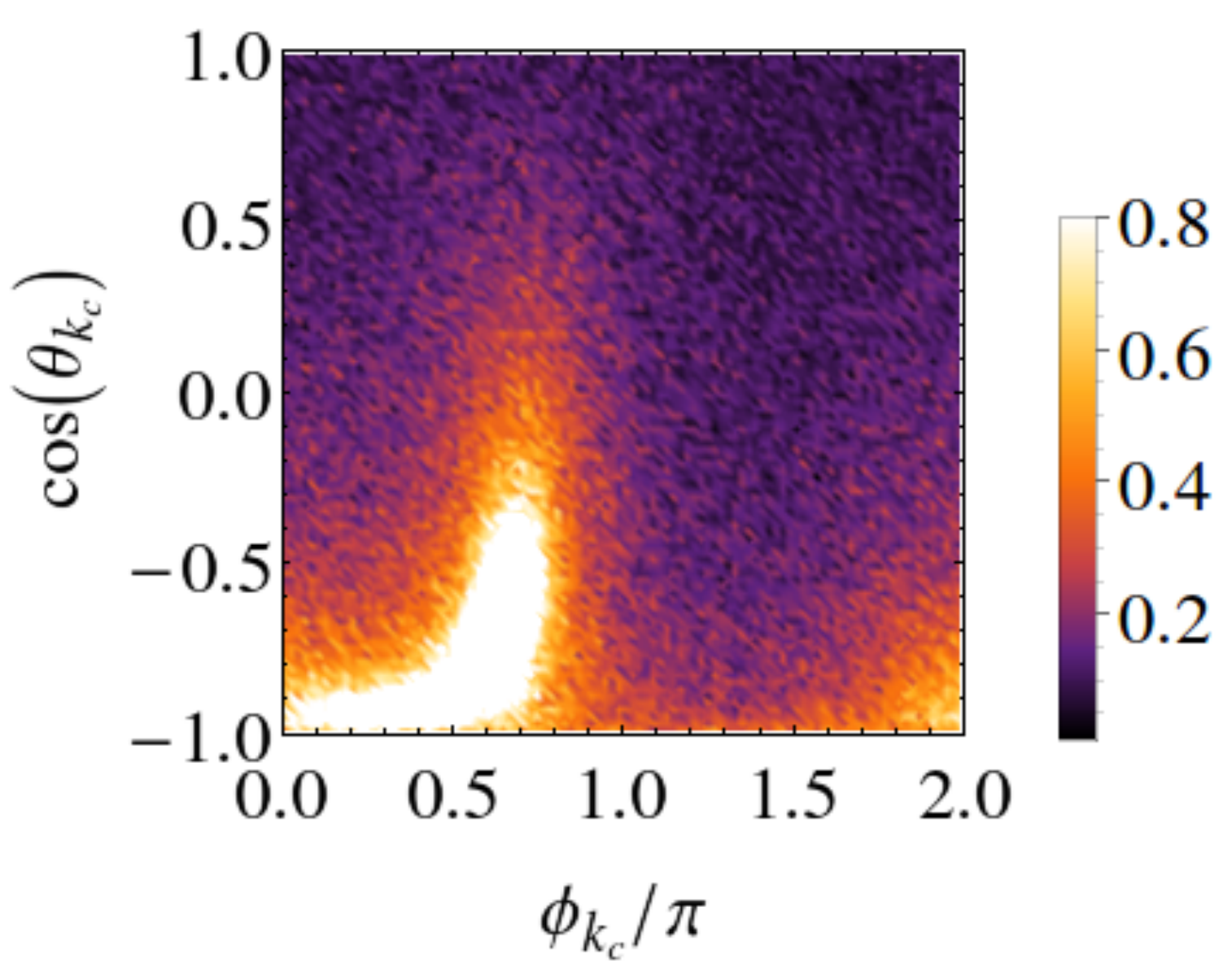}}%
{\includegraphics[width=0.49\columnwidth]{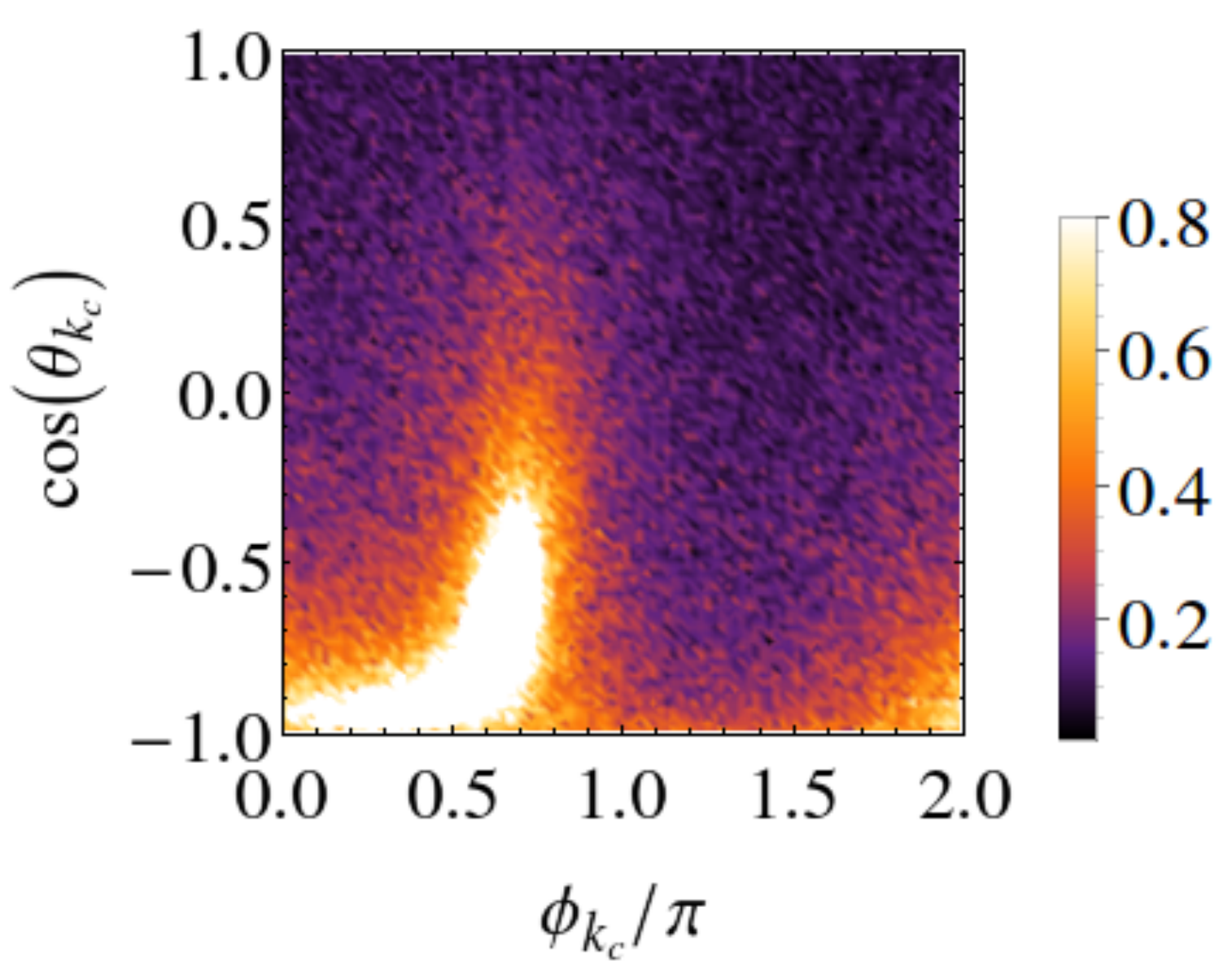}} 
\caption{The evolution of $\mathcal{P}_n(\cos \theta_{k_c},\phi_{k_c})$ as a 
function of $n=F_m$ for a coarse-grained momentum cell with $N_c=131072$ 
consecutive momenta and $k_c=19\pi/64$. The drive parameters are $g_i=4$, 
$g_f=2$, $T=2.0$ and $T=0.2$. Here $F_1=1$, $F_2=2$ and $F_m=F_{m-1}+F_{m-2}$ 
for $m>2$ with $F_m \sim \varphi^m$ for large $m$. $m$ takes the value $15$ 
(top left), $20$ (top right), $50$ (bottom left), $100$ (bottom right) in the 
four panels.} \label{fig9} \end{figure}

To better understand the NESS generated for this quasiperiodic driving 
protocol when the system is observed not stroboscopically but at $n=F_m$, we 
consider the probability distribution $\mathcal{P}_n(\cos \theta_{k_c},
\phi_{k_c})$ generated by the motion of the $N_c \gg 1$ points on the unit 
sphere for each coarse-grained momentum cell (with average momentum $k_c$) as 
a function of $n=F_m$. To do this numerically, we consider $L=16777216$ and 
divide the momentum space into $32$ cells with each coarse-grained cell having 
$262144$ consecutive momentum modes. Here we show the evolution of 
$\mathcal{P}_n(\cos \theta_{k_c},\phi_{k_c})$ as a function of $n=F_m$ 
for two drive protocols with $g_i=4$, $g_f=2$, $T=0.2$ and $T=0.05$ 
(Fig.~\ref{fig8}) and with $g_i=4$, $g_f=2$, $T=2.0$ and $T=0.2$ 
(Fig.~\ref{fig9}). The momentum cell chosen has an average momentum of $k_c = 
31\pi/64$ in the former case and $k_c = 19\pi/64$ in the latter case. In both 
the cases, the $N_c$ points start from the south pole of the unit sphere at 
$n=0$ due to the choice of $|\psi(n=0)\rangle$ taken here.

From the results presented in Fig.~\ref{fig8} and Fig.~\ref{fig9}, we clearly 
see that $\mathcal{P}_n(\cos \theta_{k_c},\phi_{k_c})$ does have a well-defined 
long time limit $\mathcal{P}_\infty(\cos \theta_{k_c},\phi_{k_c})$ when the 
system is observed at $n=F_m$. The rate of convergence to $\mathcal{P}_\infty
(\cos \theta_{k_c},\phi_{k_c})$ again depends on the details of the drive 
protocol and on $k_c$. Most importantly, $\mathcal{P}_\infty(\cos 
\theta_{k_c},\phi_{k_c})$ is neither a circle on the unit sphere as is 
expected for a p-GGE (or, a g-GGE~\cite{NandySS2017}) nor a uniform 
cover of its surface as is the case for an ITE~\cite{NandySS2017} but rather 
something intermediate as shown schematically in Fig.~\ref{fig2}. As can be 
seen from Fig.~\ref{fig8} and Fig.~\ref{fig9}, the exact form of 
$\mathcal{P}_\infty(\cos \theta_{k_c},\phi_{k_c})$ is a function of the drive 
parameters and $k_c$ which suggests that the exact form of the NESS can be 
tuned by changing these parameters appropriately. 

\subsection{KKT invariant for the Fibonacci sequence}
\label{sec:IIId}

\begin{figure}
{\includegraphics[width=\hsize]{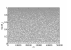}}\\
{\includegraphics[width=\hsize]{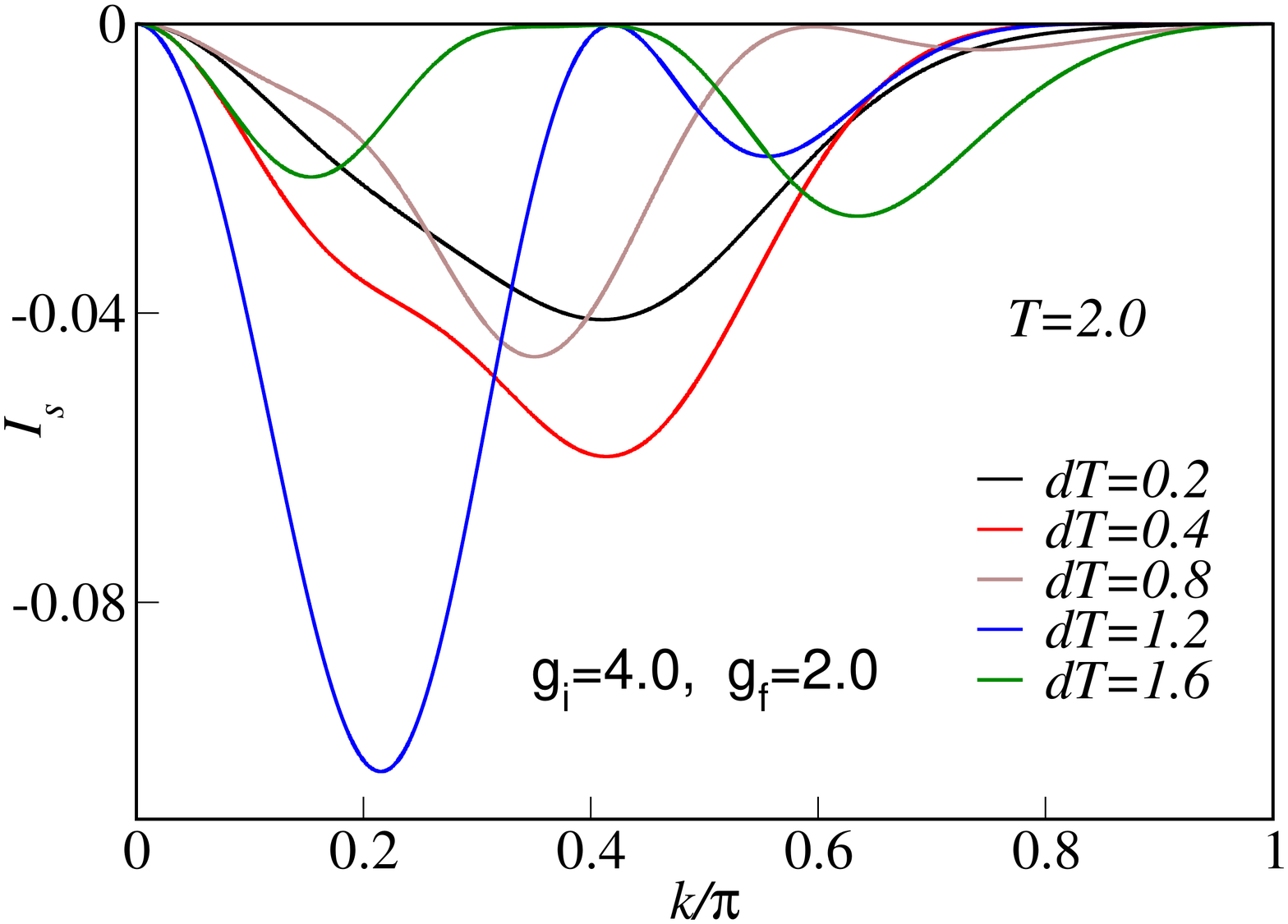}}
\caption{(Top panel) The evolution of $\sin \theta_m$ as a function of $m$ for 
the Fibonacci sequence defined in Eq.~\eqref{Fibonacci_U} for a mode with
$k= 2\pi/5$, and $g_i=4$, $g_f=2$, $T=2.0$ and $dT=1.2$. With these parameters,
$I_s \approx -0.0016877$. (Bottom panel) The KKT invariant $I_s$ as a 
function of $k$ for a fixed $g_i=4$, $g_f=2$, $T=2.0$, and different values 
of $dT$.} \label{fig10} \end{figure}

Given any two SU(2) matrices $U_1$ and $U_2$, a Fibonacci sequence of matrices 
is defined by the recursion relation shown in Eq.~\eqref{Fibonacci_U}. Let us 
parametrize $U_m$ as 
\bea U_m ~=~ e^{i\theta_m \hat{n}_m \cdot \vec{\sigma}}, \label{Um_form} \eea
where $0 \leq \theta_m \leq \pi$ and $\hat{n}_m$ is a unit vector.
We also define 
\bea \Ga_{m,m+1} ~=~ \cos^{-1} (\hat{n}_m \cdot \hat{n}_{m+1}), \label{Gamma}
\eea
where $0 \leq \Ga_{m,m+1} \leq \pi$. It was shown by Kohmoto, Kadanoff and 
Tang (Ref.~\onlinecite{KohmotoKT1983}) and Sutherland 
(Ref.~\onlinecite{Sutherland1986}) that a quantity defined as
\bea I_s ~=~ - ~(\sin \theta_m ~\sin \theta_{m+1} ~\sin \Ga_{m,m+1})^2 
\label{KKT} \eea 
is independent of $m$; we will call this the KKT invariant henceforth. 
This places a simple constraint on the allowed values 
of $\theta_m$ since from Eq.~\eqref{Gamma}, we have 
\bea \sin \theta_m ~=~ \frac{\sqrt{|I_s|}}{\sin \theta_{m+1} \sin \Ga_{m,m+1}} 
~\ge~ \sqrt{|I_s|}. \label{constraint} \eea 
In Fig.~\ref{fig10} (top panel), we show that this constraint is indeed 
satisfied in our numerics for any $k$ mode when the unitary evolution matrix 
is calculated at $n=F_m$. The KKT invariant $I_s$ is strongly dependent on $k$
and $dT$ for a fixed set of values of $g_i$, $g_f$ and $T$ as can be seen from 
Fig.~\ref{fig10} (bottom panel). The strong dependence of $I_s$ on the drive 
parameters and momentum $k$ makes the dependence of $\mathcal{P}_\infty(\cos 
\theta_{k_c},\phi_{k_c})$ on the drive parameters and $k_c$ plausible. In 
particular, if $I_s$ is close to $-1$, then the allowed values of $\theta_m$ 
are strongly constrained. We should stress here that the constraint on 
$\theta_m$ does not imply that the motion of the state $|\psi_k \rangle$ 
generated by $U_m$ necessarily has forbidden regions on the Bloch sphere 
(however, see the next section).

\section{Analysis of a toy problem}
\label{sec:IV}

In this section, instead of taking $U_1 = U_k (T+dT)$ and $U_2 = U_k(T-dT)
U_k(T+dT)$ in Eq.~\eqref{Fibonacci_U}, we will consider a simpler problem 
of a single two-level system with the following initial matrices $U_1$ 
and $U_2$,
\bea U_1 &=& e^{-i\sigma_z \left(\frac{\pi}{2}+\ep \right)}, \non \\
U_2 &=& e^{-i\sigma_x \left(\frac{\pi}{2}+\ep \right)}, \label{toyU} \eea 
and follow the motion of a state $|\psi \rangle = (0,1)^T$ on the Bloch sphere 
under the action of $U_m$ (generated by the recursion in 
Eq.~\eqref{Fibonacci_U})
as a function of $m$. For $\ep=0$, $U_m$ for any $m$ will be of the form 
$\pm e^{-i\sigma_a \left(\pi /2 \right)}$ where $a=x,y,z$. Thus the state 
$|\psi \rangle$, when represented on the Bloch sphere, will equal one of the 
eight points $(0,0,\pm 1), (0, \pm 1, 0), (\pm 1,0,0)$ as a function of $m$. 

\begin{figure}
{\includegraphics[width=0.5 \columnwidth]{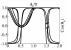}}%
{\includegraphics[width=0.5 \columnwidth]{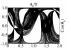}}\\
{\includegraphics[width=0.5 \columnwidth]{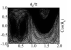}}%
{\includegraphics[width=0.5 \columnwidth]{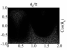}} 
\caption{The trajectory of a state $|\psi \rangle$ as a function of $m$ on the 
surface of the (unit) Bloch sphere for $m=10^6$. The Fibonacci sequence of 
matrices $U_m$ is defined by the sequence Eq.~\eqref{Fibonacci_U} and $U_1,U_2$ 
have the forms shown in Eq.~\eqref{toyU}. The values of $\ep$ used in the 
different panels are $\ep=0.10$ (top left), $\ep=0.28$ (top right), $\ep=0.29$
(bottom left) and $\ep=0.40$ (bottom right).} \label{fig11} \end{figure}

When $\ep \neq 0$, the problem is no longer analytically tractable and 
requires a numerical analysis. We show the result of such an analysis in 
Fig.~\ref{fig11} where the trajectory of the state $|\psi \rangle$ on the 
surface of the Bloch sphere is shown for four different values of $\ep$ for a 
large value of $m =10^6$. We see that for $\ep=0.10$ (top left panel, 
Fig.~\ref{fig11}) and $\ep=0.28$ (top right panel, Fig.~\ref{fig11}), there 
are still forbidden regions on the Bloch sphere just like the case of $\ep=0$. 
However, for larger values of $\ep$ like $\ep=0.29$ (bottom left panel, 
Fig.~\ref{fig11}) and $\ep=0.40$ (bottom right panel, Fig.~\ref{fig11}), the 
trajectory seems to completely fill the surface of the sphere, though in a 
highly non-uniform manner. This simple toy problem thus illustrates the 
richness of possible structures that can emerge for suitable choices of the 
matrices $U_1$ and $U_2$ which may be controlled by choosing the exact nature 
of $g_{\mathrm{ref}}(t)$ and the momentum $k$.

We now analyze the Fibonacci sequence in the limit that the KKT invariant 
$I_s \to -1$ to understand some features of the above problem when $\ep$ is 
small. To simplify the notation, let us start at some value of $m$ and define
\bea \ta_m &\equiv& \frac{\pi}{2} ~+~ \ep_1, \non \\
\ta_{m+1} &\equiv& \frac{\pi}{2} ~+~ \ep_2, \non \\
\Ga_{m,m+1} &\equiv& \frac{\pi}{2} ~+~ \ga, 
\label{ta1} \eea
where $\ta_m$ and $\Ga_{m,m+1}$ are defined in Eqs.~(\ref{Um_form}-\ref{Gamma}).
Then $I_s$ close to $-1$ implies that $\ep_1$, $\ep_2$ and $\ga$ all lie 
close to zero. An alternate way of writing Eq.~\eqref{ta1} is 
\bea \sin \ep_1 &=& - ~\frac{1}{2} \mathrm{tr}(U_m), \non \\
\sin \ep_2 &=& - ~\frac{1}{2} \mathrm{tr}(U_{m+1}), \non \\
\sin \ga &=& - ~\hat{n}_m \cdot \hat{n}_{m+1}. \label{ta2} \eea
The KKT invariant is then given by 
\beq I_s ~=~ - 1 ~+~ \ep_1^2 ~+~ \ep_2^2 ~+~ \ga^2 \label{is2} \eeq
plus terms of fourth order in $\ep_1, ~\ep_2, ~\ga$. 
We then find from the recursion relation in Eq.~\eqref{Fibonacci_U} that, to 
third order in the small parameters $\ep_1, ~\ep_2, ~\ga$, the quantities 
defined in Eq.~\eqref{ta1} for $m$ transform as follows when $m \to m+1$:
\bea \ep_1 &\to& \ep_2, \non \\
\ep_2 &\to& - ~\ga ~-~ \ep_1 \ep_2 ~+~ \frac{1}{2} ~\ga ~(\ep_1^2 ~+~ 
\ep_2^2), \non \\
\ga &\to& \ep_1 ~-~ \ep_2 \ga ~+~ \frac{1}{2} ~\ep_1 ~(\ga^2 ~-~ \ep_2^2). 
\label{iter2} \eea
Iterating Eqs.~\eqref{iter2} six times, we find that when $ m \to m+6$,
\bea \ep_1 &\to& \ep_1 ~+~ 4 \ep_1 ~(\ga^2 - \ep_2^2), \non \\
\ep_2 &\to& \ep_2 ~+~ 4 \ep_2 ~(\ep_1^2 - \ga^2), \non \\
\ga &\to& \ga ~+~ 4 \ga ~(\ep_2^2 - \ep_1^2). \label{iter3} \eea

Since $\ep_1, ~\ep_2, ~\ga$ are small, the changes in these parameters 
given in Eqs.~\eqref{iter3} are small. It is then convenient to replace those 
equations by differential equations where we define 
\beq \frac{dz}{dm} ~\equiv~ \frac{z(m+6) ~-~ z(m)}{6}, \eeq 
where $z$ can denote any of the variables given in 
Eqs.~\eqref{iter3}. We then obtain the equations
\bea \frac{d\ep_1}{dm} &=& \frac{2}{3}~ \ep_1 ~(\ga^2 - \ep_2^2), \non \\
\frac{d\ep_2}{dm} &=& \frac{2}{3}~ \ep_2 ~(\ep_1^2 - \ga^2), \non \\
\frac{d\ga}{dm} &=& \frac{2}{3}~ \ga ~(\ep_2^2 - \ep_1^2). \label{diff1} \eea

\begin{figure}
{\includegraphics[width=\hsize]{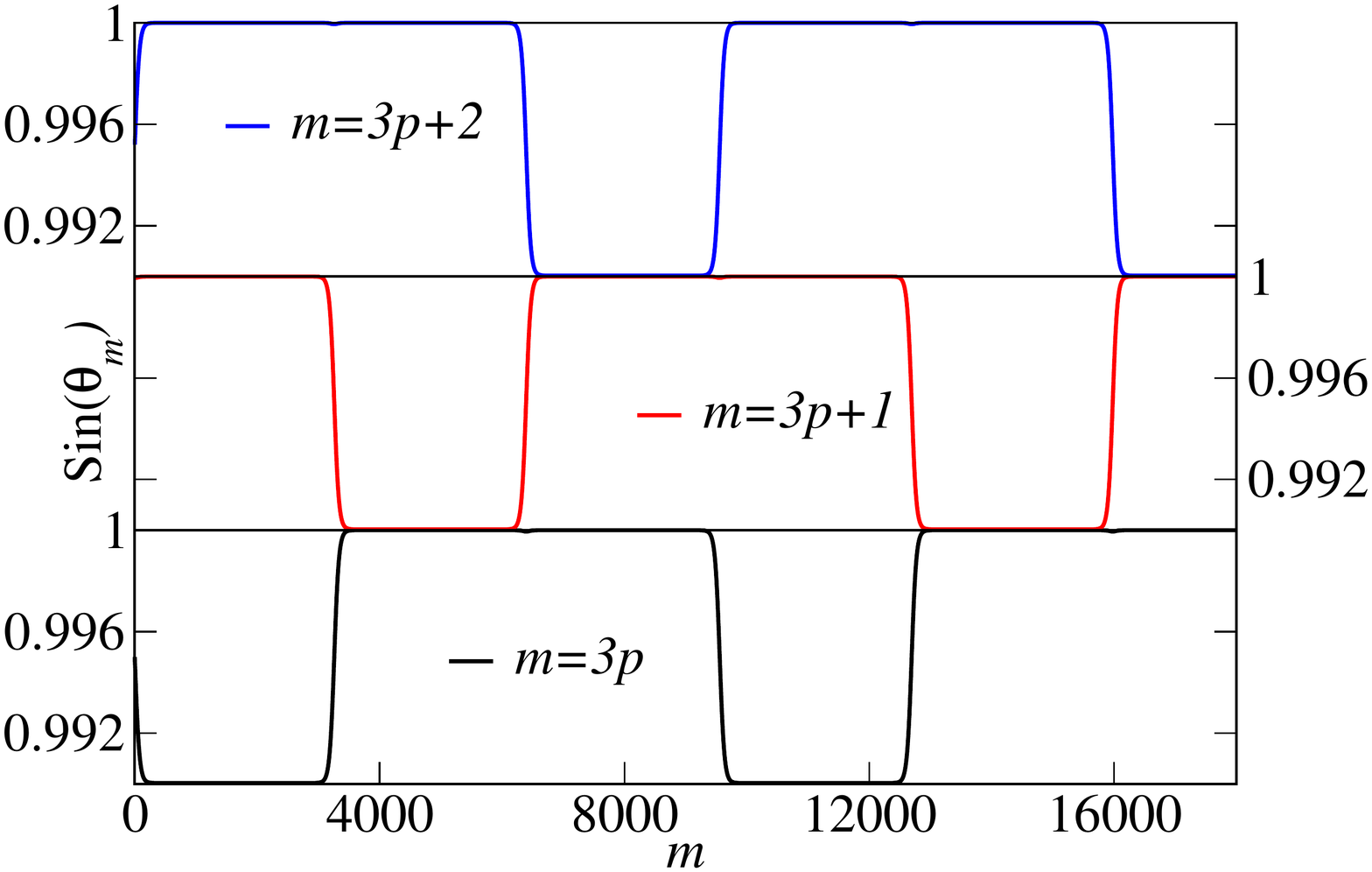}} \\
{\includegraphics[width=\hsize]{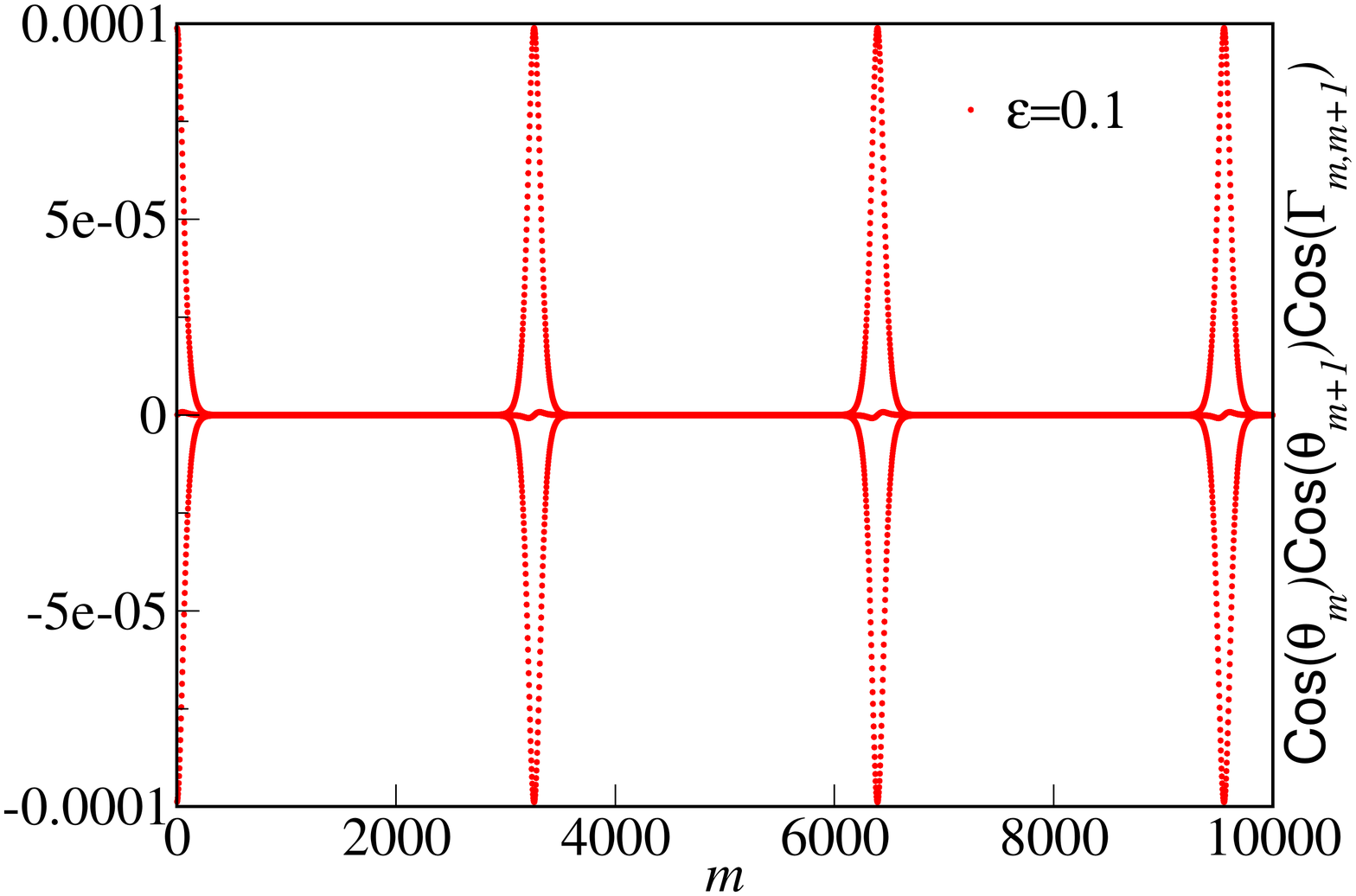}}
\caption{(Top panel) The evolution of $\sin \theta_m$ as a function of $m$ for
$\ep=0.1$. (Bottom panel) The evolution of $\cos \theta_m \cos \theta_{m+1} 
\cos \Ga_{m,m+1}$ as a function of $m$ for $\ep=0.1$.} 
 \label{fig12} 
\end{figure}

We observe that Eqs.~\eqref{diff1} conserve the invariant $I_s$ given in
Eq.~\eqref{is2}; this implies that a point with the coordinates $(\ep_1,\ep_2,
\ga)$ moves on the surface of a sphere. Interestingly, Eqs.~\eqref{diff1} 
conserve another quantity $C$ given by
\beq C ~=~ \ep_1 \ep_2 \ga. \label{c} \eeq
The presence of this invariant implies that the point $(\ep_1,\ep_2,\ga)$ moves
on closed curves on the surface of the sphere, so that we have effectively
only one quantity which changes with $m$. We thus have an integrable system to 
this order in $\ep_1, ~\ep_2, ~\ga$. [We note from Eq.~\eqref{ta1} that
$C = - \cos (\theta_m) \cos (\theta_{m+1}) \cos (\Gamma_{m,m+1})$ to third
order in $\ep_1, ~\ep_2, ~\ga$. We will plot the latter quantity in 
Fig.~\ref{fig12} below since it is easier to calculate numerically].

We will now apply the above analysis to the case defined in Eq.~\eqref{toyU}. 
The initial conditions are then given by 
\bea \ep_1 &=& \ep_2 ~=~ \ep, ~~~~\ga ~=~ 0.
\label{sourav} \eea
Eqs.~\eqref{diff1} then simplify to 
\bea \frac{d\ep_1}{dm} &=& - \frac{2}{3}~ \ep_1 \ep_2^2, \non \\
\frac{d\ep_2}{dm} &=& \frac{2}{3}~ \ep_1^2 \ep_2. \label{diff3} \eea
Since $\ep_1^2 + \ep_2^2 = 2 \ep^2$ is an invariant, we can parametrize
\beq \ep_1 ~=~ \sqrt{2} \ep \sin \phi ~~~ {\rm and} ~~~\ep_2 ~=~ \sqrt{2} 
\ep \cos \phi. \eeq
Eqs.~\eqref{diff3} imply that
\beq \frac{d\phi}{dm} ~=~ - \frac{2}{3}~ \ep^2 \sin \phi \cos \phi. 
\label{phi} \eeq
This equation has stable fixed points at $\phi = 0$ and $\pi$, and unstable
fixed points at $\phi = \pi/2$ and $3\pi/2$. At $m=0$, the initial condition 
(Eq.~\eqref{sourav}) fixes $\phi=\pi/4$. Eq.~\eqref{phi} then implies that
$\phi$ will flow to zero as $m$ increases, so that $\ep_1 \to 0$ and
$\ep_2 \to \sqrt{2} \ep$. According to Eq.~\eqref{phi}, the time scale 
(here we are thinking of $m$ as time) of approaching the fixed point should 
be of the order of $1/\ep^2$. According to the first order terms in
Eq.~\eqref{iter2}, $\ep_1$ cycles
as $\ep_1 \to \ep_2 \to - \ga \to - \ep_1 \to - \ep_2 \to \ga \to \ep_1$ for 
six successive iterations. Hence $\sin \ta_m = \cos \ep_1$ should cycle over
three different values given by $1 - \ep_1^2/2, ~1 - \ep_2^2/2, ~1 - \ga^2/2$.
Monitoring $\sin \theta_m = \cos \ep_1 \approx 1 - \ep_1^2/2$ for 
$\ep=0.1$ as a function of $m$ (Fig.~\ref{fig12}) clearly shows that 
$\ep_1, \ep_2, \ga$ do not remain at one particular fixed point
permanently; after long intervals of time, they move from one fixed point to
another relatively quickly. For example, for $\ep=0.1$, this movement
happens at intervals of about $m=3200$; the movement itself occurs over a
duration of $m=100$ (which is equal to $1/\ep^2$). We will argue below 
that these movements may be understood qualitatively by appealing to terms
higher than third order which we have ignored when deriving
Eqs.~\eqref{diff1}.

We will begin by finding the fixed points of Eqs.~\eqref{diff1}. The KKT 
invariant implies that the three parameters lie on the surface of a sphere 
\beq \ep_1^2 ~+~ \ep_2^2 ~+~ \ga^2 ~=~ r^2, \label{sph} \eeq
where $r$ is a small number. We then find that there are two types of 
fixed points. \\ \\
\noi (i) Any two of the parameters $\ep_1, ~\ep_2, ~\ga$ are equal to zero, 
and the third one is equal to $\pm r$. This gives six possible fixed 
points. \\ \\
\noi (ii) $\ep_1^2 = \ep_2^2 = \ga^2 = r^2/3$. This gives eight possible
fixed points. \\ \\
Next, we look at the stability of the above fixed points. Denoting the 
first order deviations of the three parameters from a fixed point by $\de_i$ 
(where $i=1,2,3$), we obtain equations of the form
\beq \frac{d \de_i}{dm} ~=~ \sum_{j=1}^3 ~M_{ij} \de_j. \eeq
The eigenvalues of the matrix $M$ determine the stability of a fixed point; 
if any one of the eigenvalues has a positive real part, the fixed point is 
unstable. For the six fixed points of type (i), the eigenvalues turn out to 
be 0 and $\pm (2/3) r^2$, implying that each of these fixed points is 
unstable along some direction. For the eight fixed points of type 
(ii), the eigenvalues are 0 and $\pm i (2/3\sqrt{3}) r^2$; hence a small
deviation from these fixed points will oscillate but not grow with time.

The invariant $C = \ep_1 \ep_2 \ga$ defined in Eq.~\eqref{c} is equal to 0 
for the fixed points of type (i) and $\pm r^3 /(3\sqrt{3})$ for the fixed 
points of type (ii). 
For the cases studied numerically, i.e., $(\ep_1, 
\ep_2,\ga)=(\ep,\ep,0)$ at $m=0$, we have $r = \sqrt{2} \ep$ and $C=0$. 
For this case, we have seen above that the systems flows to the fixed
point $(0,\sqrt{2} \ep,0)$ over a time scale of order $1/\ep^2$.
However, this is an unstable fixed point; hence even a small deviation from
this fixed point in an unstable direction will grow exponentially with $m$.

We now note that while the KKT invariant is an exact invariant, 
the quantity $C$ defined in Eq.~\eqref{c} is invariant only up to terms 
of third order in $\ep_1, ~\ep_2, ~\ga$. Considering the possible effects of 
terms of higher than third order on the right hand sides of Eqs.~\eqref{diff1},
we may expect that over very long periods of time (i.e., periods which are 
much larger than $1/\ep^2$), the three parameters will differ from what 
we would obtain if only the third order terms were present in 
Eqs.~\eqref{diff1}. As a result, we would also expect $C$ to deviate 
eventually from its initial value, i.e., from $C=0$. We therefore 
expect that over long times, $|C|$ will not exactly be equal to zero, although
it will remain much smaller than the largest possible value of $r^3/(3 
\sqrt{3}) \simeq 0.00054$. Fig.~\ref{fig12} confirms this numerically for 
the case $\ep=0.1$ where we monitor $\cos \theta_m \cos \theta_{m+1} \cos
\Ga_{m,m+1}$ which is equal to $-C$ for small $\ep_1, \ep_2, \ga$.

Given that $C$ is very small but not exactly equal to zero, we can now 
see how the parameters $\ep_1, ~\ep_2, ~\ga$ will change over very long periods
of time. In the beginning, when $\ga$ is very close to zero, we have seen that
$\ep_1$ flows to zero while $\ep_2$ flows to $\sqrt{2} \ep$; these
flows occur over a time scale of order $1/\ep^2 = 100$. According to
Eqs.~\eqref{diff1}, this is a stable fixed point for $\ep_1$ but not for $\ga$.
Eventually, therefore, $\ga$ will start growing while $\ep_1$ will remain
small. Once $\ga$ becomes larger than $\ep_1$, $\ep_2$ will start
becoming smaller. Thus the behaviors of $\ep_1, ~\ep_2, ~\ga$ will get
cyclically interchanged; namely, $\ep_1$ will stay very close to zero, 
$\ep_2$ will flow to zero and $\ga$ will flow to $\sqrt{2} \ep$. Eventually,
this behavior again gets interchanged; $\ep_2$ stays very close to zero while 
$\ga$ and $\ep_1$ flow to zero and $\sqrt{2} \ep$ respectively. This is 
indicated in Fig.~\ref{fig12} where we see that out of the three quantities
$1 - \ep_1^2/2$, $1 - \ep_2^2/2$, and $1 - \ga^2/2$, two of them remain
close to one while the third one remains close to $1 - \ep^2 = 0.99$ at all 
times (except for some rapid changes occurring over a time scale of order 
$1/\ep^2$). We see that this behavior gets cyclically interchanged after time 
intervals of the order of 3200. We cannot quantitatively explain the value 
3200 since we do not know the form of the higher order terms that will appear 
on the right hand sides of Eqs.~\eqref{diff1}; however, as argued above, we 
do expect this time scale to be much larger than $1/\ep^2 = 100$.

\section{Conclusions and outlook}
\label{sec:V}

In this work, we have considered a prototypical many-body integrable model, the
one-dimensional transverse field Ising model, that is continually driven 
by a time-dependent magnetic field resulting in a unitary dynamics. The 
translational symmetry of the system allows for a mapping to a pseudospin 
representation for each momentum $k$, where the pseudospin is driven by
its own time-dependent pseudo-magnetic field that varies with 
$k$ (Eq.~\eqref{pseudospinH}). In the thermodynamic limit, for a subsystem 
whose size is much smaller than that of the total system, the reduced density
matrix of the subsystem is fixed by certain coarse-grained quantities in
momentum space (Eq.~\eqref{coarsegrain}). These coarse-grained quantities have 
a well-defined large $n$ limit in case a nonequilibrium steady state exists at
late times. We also define a probability distribution on the unit sphere for 
each value of average $k_c$ by using a large number of consecutive momenta in 
a small
cell that is centered around $k_c$ and using the pseudospin representation for
each $k$ to represent the corresponding state on the unit sphere. This
probability distribution for each $k_c$ changes in an irreversible manner and
approaches a steady state distribution in the limit $n \to \infty$. 

We consider a driving protocol with two types of square pulses with possible
periods $T+dT$ and $T-dT$ such that the unitary time evolution operators 
$U_k(T+dT)$ and $U_k(T-dT)$ do not commute with each other. These pulses
alternate according to a Fibonacci sequence to mimic a quasiperiodic drive
that is neither periodic nor random in time. We find that neither the local
quantities nor the coarse-grained quantities approach a well-defined steady
state if the system is viewed stroboscopically for even $n \sim
10^5$. However, when the system is viewed at $n=F_m$ (where $F_m$ are the
Fibonacci numbers which increase as $\varphi^m$ for large $m$), a well-defined
nonequilibrium steady state indeed emerges at exponentially late times. This
exponentially late approach to the steady state was also previously found by
us for another driving protocol that was neither random nor 
periodic~\cite{NandySS2017}. We believe that this may be a generic feature
of quasiperiodically driven integrable systems.

To characterize the nonequilibrium steady state, we also look at the
probability distribution of consecutive momenta in a momentum cell centered at
$k_c$. We find that the final probability distribution is qualitatively
different compared to the distributions that arise for a periodically driven
system or a system that locally heats up to an infinite temperature ensemble,
i.e., it is neither a circle nor a uniform covering of the unit sphere. 
Furthermore, the form of the distribution is sensitive to the parameters of 
the driving protocol and the value of $k_c$. We also study a toy problem 
of a single two-level system where the unitary matrices are arranged in a 
Fibonacci sequence to illustrate how the nature of the distribution on the 
Bloch sphere is highly sensitive to the parameters of the problem. In 
the limit where the KKT invariant is close to $-1$, the problem can be studied 
analytically to a large extent, and we find an intricate pattern for the state 
at late times.

A deeper understanding of these distributions and their tunability with the 
drive parameters and $k_c$ is highly desirable since that would lead to a wide 
variety of nonequilibrium steady states that were previously unknown. Another
open question is to come up with a simple ensemble description for the 
resulting steady state since it lies beyond a periodic generalized Gibbs
ensemble approach. Finally, we note that the one-dimensional transverse 
field Ising model driven by a Fibonacci sequence of square pulses has been 
recently studied in the high frequency limit, and it has been shown that a
nonequilibrium steady state emerges in that limit which resembles the steady
state of a periodically driven system~\cite{Maity2018}.

\section*{Acknowledgements}
A.S. is grateful to Sthitadhi Roy for useful discussions. The work of A.S. 
is partly supported through the Partner Group program between the Indian 
Association for the Cultivation of Science (Kolkata) and the Max Planck 
Institute for the Physics of Complex Systems (Dresden). D.S. thanks Department 
of Science and Technology, India for Project No. SR/S2/JCB-44/2010 for 
financial support.


\end{document}